\documentclass[a4paper,fleqn]{cas-sc}

\usepackage[numbers]{natbib}
\usepackage{graphicx}
\usepackage{amssymb}
\usepackage{amsmath}
\usepackage[ruled,vlined]{algorithm2e}
\usepackage[section]{placeins}

\begin{document}
\let\WriteBookmarks\relax
\def\floatpagepagefraction{1}
\def\textpagefraction{.001}

\shorttitle{Joint Voxel Flow-Phase Field Framework}
\shortauthors{Zhou et al.}

\title[mode = title]{A Joint Voxel Flow-Phase Field Framework for Ultra-Long Microstructure Evolution Prediction with Physical Regularization}

\author[1]{Ao Zhou}
\credit{Data Curation, Formal Analysis, Investigation, Methodology, Software, Visualization,Writing - Original Draft}
\author[1]{Salma Zahran}
\credit{Methodology}
\author[1]{Chi Chen}
\credit{Methodology}
\author[1,2]{Zhengyang Zhang}
\cormark[1]
\ead{zhangzhengyang3@xiaomi.com}
\credit{Investigation, Methodology, Software}
\author[3]{Yanming Wang}[orcid=0000-0002-0912-681X]
\cormark[2]
\ead{yanming.wang@sjtu.edu.cn}
\ead[URL]{https://sites.gc.sjtu.edu.cn/yanming-wang/}
\credit{Conceptualization, Funding Acquisition, Supervision}

\affiliation[1]{organization={Global College, Shanghai Jiao Tong University},
            city={Shanghai},
            postcode={200240},
            country={China}}
\affiliation[2]{organization={Xiaomi Company},
            city={Beijing},
            country={China}}
\affiliation[3]{organization={Global Institute of Future Technology, Shanghai Jiao Tong University},
            city={Shanghai},
            postcode={200240},
            country={China}}

\cortext[1]{Corresponding author.}
\cortext[2]{Principal Corresponding author.}

\begin{abstract}
Phase-field (PF) modeling is a powerful tool for simulating microstructure evolution. To accelerate the simulation of PF models governed by complex PDEs, machine learning methods such as PINNs and ConvLSTM have been introduced. However, current machine-learning-based approaches still suffer from limited flexibility, poor generalization, and short prediction horizons. To address these challenges, we present a joint framework that couples a voxel-flow network (VFN) with PF simulations in an alternating manner for long-horizon prediction of microstructure evolution with substantial computational acceleration. The VFN iteratively predicts future evolution by generating the next snapshot from the previous two snapshots. Periodic PF simulations suppress nonphysical artifacts, reduce accumulated error, and extend the reliable prediction horizon. The VFN was validated using a grain-growth example, and its accuracy outperforms that of similar prediction methods while preserving topological grain details. For an ultra-long grain-growth prediction of 82 frames from 2 input frames, the grain number decreases from 600 to 29 while the NMSE of the average grain area remains 1.64\%. The framework also exhibits good generalizability across different PF models. Overall, this joint framework enables rapid, flexible, generalizable, and physically consistent microstructure forecasting from image-based data over ultra-long time scales.
\end{abstract}

\begin{graphicalabstract}
\includegraphics[width=\linewidth]{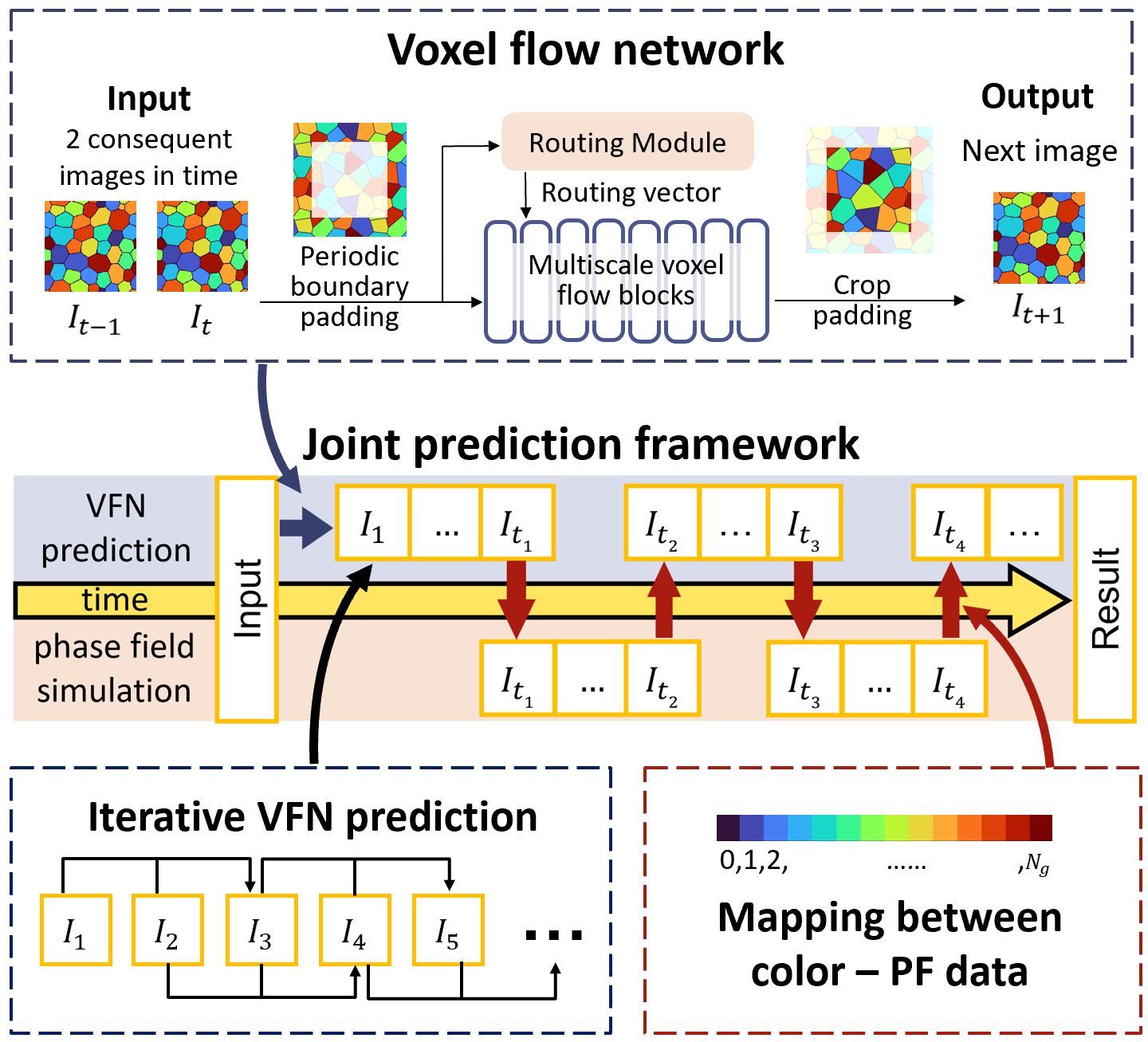}
\end{graphicalabstract}

\begin{highlights}
\item A joint framework combines VFN and phase-field for evolution prediction.
\item Physical regularization in the framework controls error of prediction results.
\item The framework enables stable and ultra-long microstructure evolution prediction.
\item The method was validated on grain growth and spinodal decomposition benchmarks.
\item High accuracy, strong generalization, and notable computational speedup are achieved.
\end{highlights}


\maketitle

\section{Introduction}
\label{sec:Introduction}

Phase-field models \cite{RN95} are widely used to simulate microstructure evolution in a broad range of materials problems, including sintering \cite{RN92}, grain growth \cite{chen1994computer}, dendrite solidification and deposition \cite{RN93,RN94}, and spinodal decomposition \cite{cahn1961spinodal}. Governed by partial differential equations (PDEs), these models provide a powerful framework for describing the complex dynamics of phase transformations. However, their high computational cost often limits their use, especially for long-time evolution or large simulation domains \cite{RN97}. To overcome this limitation, recent studies have introduced machine learning (ML) methods to accelerate phase-field simulations \cite{RN96, wang2022multi}. Existing approaches can be broadly divided into two categories: physics-informed methods, represented by physics-informed neural networks (PINNs) \cite{PINNoriginal}, which attempt to solve the governing equations more efficiently; and purely data-driven methods, which learn the evolution directly from simulation or experimental data without explicitly solving the PDEs \cite{lanzoni2024extreme}. The former emphasizes physical consistency, whereas the latter prioritizes flexibility and computational efficiency.

Physics-informed neural networks incorporate physical laws directly into the training process by embedding PDE residuals into the loss function \cite{PINNoriginal}. In phase-field problems, this typically means optimizing not only the mismatch between prediction and reference data, but also the residuals of governing equations such as the Allen--Cahn and Cahn--Hilliard equations \cite{wight2020solving, Haridasan2025}. To improve accuracy and convergence, many variants have been proposed, including hard enforcement of initial and boundary conditions \cite{pinnhardconstraint}, adaptive sampling and NTK-based loss weighting in PF-PINNs \cite{pfpinnschen}, and time-marching PINN formulations for coupled Cahn--Hilliard--Navier--Stokes systems \cite{pinnphaseflow}. Related developments include PINNs-MPF for multi-phase problems \cite{pinnsmpf}, as well as physics-informed neural operators such as DeepONet \cite{deeponetbazant} and Fourier-domain PINOs \cite{gangmei2025}. Despite these advances, physics-informed approaches still face important limitations. They often require careful loss balancing, extensive training, and explicit prior knowledge of the governing equations, which reduces their efficiency, scalability, and flexibility for long-time or large-scale microstructure evolution. They are also less suitable for image-based or experimental applications in which no explicit phase-field formulation is available. Moreover, because PINNs usually require problem-specific architectures and customized loss functions, their generalizability across different phase-field systems remains limited. This difficulty becomes even more severe for complex multi-phase-field models. For example, the grain-growth phase-field model involves multiple order parameters, each governed by a separate Allen--Cahn equation \cite{chen1994computer}, which would require a highly complicated composite loss function to enforce all coupled dynamics simultaneously. As a result, PINNs have not yet become a practical solution for grain-growth prediction.

In contrast, data-driven approaches learn microstructure evolution directly from data using architectures such as convolutional neural networks (CNNs), recurrent neural networks (RNNs), long short-term memory networks (LSTMs), and graph neural networks (GNNs). Convolutional recurrent neural networks have been used to predict spinodal decomposition, including three-dimensional evolution with physics-inspired designs for long-term extrapolation \cite{lanzoni2024extreme}. PredRNN-style models based on recurrent architectures have also been applied to microstructure prediction \cite{farizhandi2023spatiotemporal}. CNN-based frameworks combined with recurrent modules have been used for dendritic growth, grain growth, and spinodal decomposition \cite{zhu2024spatiotemporal, yang2021self}, while graph neural networks have shown superior performance for grain growth by explicitly encoding topological relationships \cite{fan2024accelerate}. GrainGNN further extended this idea to 3D epitaxial grain growth \cite{qin2024graingnn}. Other recent studies combined graph convolutional networks, LSTMs, and physical constraints to predict 2D and 3D spinodal decomposition with good accuracy and generalizability \cite{razavi2026physics}. Additional efforts have explored low-dimensional latent representations with RNNs, LSTMs \cite{Zapiain2021accelerating, Hu2022accelerating}, and deep neural operators \cite{oommen2022learning}, while architectures such as ResNet, U-Net, and Y-Net have also been adopted for various phase-field prediction tasks \cite{ciesielski2025deep, tseng2023deep, chen2024mau, rieger2024setting, peivaste2022machine, yan2022novel}. Nevertheless, most existing data-driven models still struggle to maintain accuracy over prediction horizons longer than those represented in the training data, which limits their practical use. In addition, grain-growth studies often rely on binary-colored images, making it difficult to distinguish individual grains and reducing the effectiveness of downstream quantitative analysis.

Because microstructure evolution can also be regarded as a temporal sequence of images, it is natural to adapt advanced video prediction models to phase-field evolution. One recent example used the SimVP video prediction model \cite{RN99} combined with multi-order aggregation features to build a model called "SimGate" \cite{wu2025simgate} for predicting the sintering of polycrystalline particles. Another important class of video prediction methods is based on optical flow, which predicts future frames by estimating pixel motion between neighboring frames, thereby achieving lightweight yet powerful predictive capability. The voxel-flow network \cite{hu2023dynamic} is an optical-flow-based method that predicts the next video frame from two previous frames. It provides end-to-end prediction by estimating per-pixel motion between adjacent frames and dynamically adjusting a scaling factor to capture motion at different scales while preserving correct spatial information. Because the time evolution of phase-field models also contains motion across multiple scales, integrating VFN with phase-field prediction is a promising strategy.

Considering the challenges of current prediction methods and the potential benefit of VFN, we choose to create a joint prediction framework that combines VFN and physical error correction for longer horizon, generalized phase field predictions. The VFN is used for rapid prediction of microstructure evolution, while short phase-field simulations are periodically inserted to correct accumulated prediction errors. Unlike traditional methods that solve PDEs or extract intermediate features, our approach treats evolving microstructures as a sequence of video frames, directly learning the spatiotemporal evolution from the pixel-level dynamics of phase-field images. We adopt colored phase-field configuration snapshots, in contrast to previous studies that rely on binary coloring, which facilitates seamless transfer between VFN prediction and phase-field simulation. By leveraging high-fidelity phase-field simulations as training data, the VFN captures both local and global features of microstructure evolution with high accuracy. In the following sections, we will demonstrate the VFN prediction of grain growth and spinodal decomposition process, then use joint framework to predict long time scale grain growth evolution to show its robust ability to do prediction on various and long time range microstructure evolution.

\section{Methods}
\label{sec:Methods}

\begin{figure}[pos=htbp]
    \centering
    \includegraphics[width=0.7\linewidth]{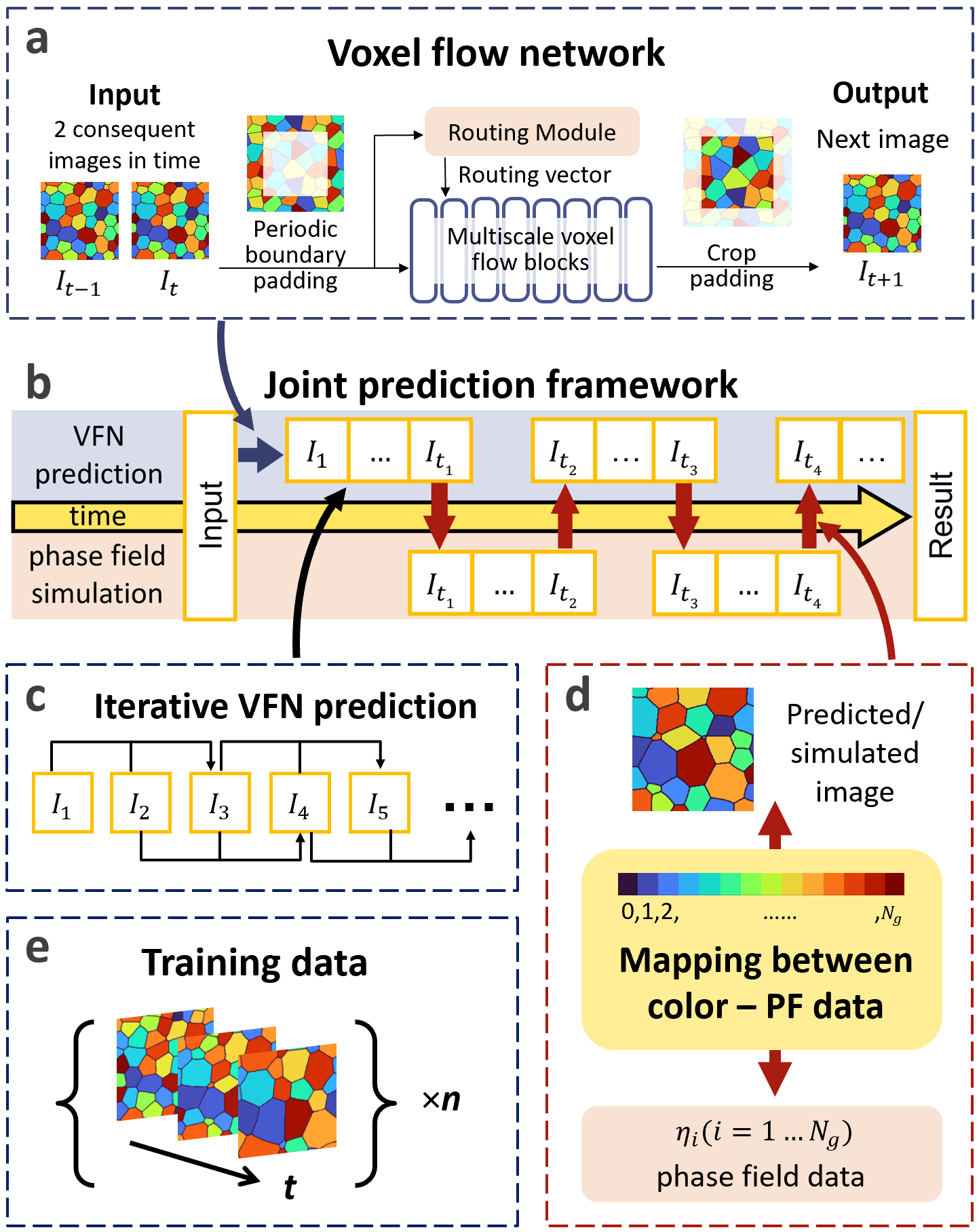}
    \caption{Diagram of the joint prediction framework. (a) Voxel-flow network with a padding--cropping operation for enforcing periodic boundaries. (b) Joint prediction framework integrating the voxel-flow network with corrective phase-field simulations. (c) Iterative VFN prediction procedure, in which each predicted output frame is fed back as the next input. (d) Image--data interface based on colormap mapping between RGB pixels and phase-field variables. (e) Phase-field evolution image datasets used for training, validation, and testing.}
    \label{fig:method}
\end{figure}

\subsection{Joint prediction framework}
\label{sec:framework}

By treating microstructure evolution as a sequence of images, we formulate phase-field dynamics as a video prediction problem and develop a joint framework that couples data-driven voxel-flow prediction with physics-based phase-field (PF) simulations. As illustrated in figure \ref{fig:method}(b), the voxel-flow network (VFN) and PF solver are alternately employed to predict long-term microstructure evolution. The VFN predicts successive microstructural frames directly from image inputs, while short PF simulation segments are periodically inserted to correct accumulated errors and maintain physical consistency. The linkage between images and phase-field variables is enabled by an image--data mapping procedure (figure \ref{fig:method}(d)), which is described in detail below.

The joint prediction proceeds as follows. Starting from two consecutive images, the VFN iteratively predicts several future frames $I_1 \rightarrow I_{t_1}$. The phase-field configuration is then reconstructed from the final predicted image via the image--data mapping and used as the initial condition for a short PF simulation, producing frames $I_{t_1} \rightarrow I_{t_2}$. The last two PF snapshots are subsequently fed back into the VFN, which resumes prediction for the next segment $I_{t_2} \rightarrow I_{t_3}$. This alternating procedure can be repeated multiple times to reach arbitrarily long prediction horizons $I_{t_4}, I_{t_5}, \ldots$. Because the time span available in the training data is finite, VFN-only prediction tends to accumulate errors when extrapolated far beyond the training range. By periodically inserting short corrective PF simulations, the joint framework effectively suppresses error growth and substantially extends the feasible prediction horizon.

The VFN architecture, shown in figure \ref{fig:method}(a), consists of nine sequential voxel flow blocks (VFBs; Section~\ref{sec:VFB}) operating at progressively finer effective resolutions, together with a routing module (Section~\ref{sec:VFN}) that computes a routing vector for each prediction. Prior to VFN inference, input PF snapshot images are padded to enforce periodic boundary conditions. After prediction, the padded boundaries are discarded and only the central region is retained. The routing vector dynamically determines which VFB scales are activated based on the estimated motion magnitude between the two input frames~\cite{hu2023dynamic}. Each VFB predicts incremental refinements to both the voxel flow and the blending mask, while the routing mechanism adaptively selects the relevant refinement stages during training and inference. This design enables efficient allocation of computational resources by executing only the necessary blocks for a given input while maintaining high predictive accuracy.

Given two consecutive input frames $I_{t-1}$ and $I_t$, the VFN predicts the next frame $I_{t+1}$, which is then recursively fed back as input to generate longer trajectories (figure \ref{fig:method}(c)). Specifically, the network estimates backward optical flows $f_{t+1 \rightarrow t}$ and $f_{t+1 \rightarrow t-1}$, which map pixels from the predicted frame to the two preceding frames. Using pixel-wise backward warping, denoted by $\overleftarrow{W}(\cdot)$, and a fusion mask $m$ that accounts for discontinuous motion such as occlusion or newly exposed regions, the predicted frame is reconstructed as
\begin{equation}
\label{eq:It+1}
I_{t+1}=m \times \overleftarrow{W}(I_{t-1}, f_{t+1 \to t-1})+(1 - m) \times \overleftarrow{W}(I_t, f_{t+1 \to t}).
\end{equation}
For notational convenience, we denote the collection of optical flows and the fusion mask as the voxel flow $F_{t+1}$. Equation~(\ref{eq:It+1}) can then be compactly written as
\begin{equation}
I_{t+1} = \overleftarrow{W}(I_{t-1}, I_t, F_{t+1}),
\end{equation}
indicating that predicting $I_{t+1}$ reduces to estimating the voxel flow $F_{t+1}$.

To enable joint simulation and quantitative analysis, we introduce a forward--reverse colormap mapping between images and underlying phase-field configurations, as illustrated in figure \ref{fig:method}(d). For the grain growth model, each color in the turbo colormap corresponds to a distinct grain order parameter $\eta_i \in [0,1]$, $i = 1,2,\ldots,N_g$, where $N_g$ is the total number of grains. In the forward mapping, a grain index matrix with values $\{0,1,2,\ldots,N_g\}$ is constructed, where 0 denotes grain boundaries, and each index is mapped to an RGB value using the colormap. To reconstruct phase-field data from predicted images, a reverse mapping is performed: for each pixel, its RGB value is compared with the colormap using the Euclidean distance in RGB space, and the closest match $i$ is selected to recover $\eta_i = 1$ and $\eta_j = 0$ for $j \neq i$. Pixels identified as grain boundaries (index 0) are assigned $\eta_i = 0$ for all order parameters $i = 1,2,\ldots, N_g$. Because this reverse mapping assigns each pixel to a single grain index, the reconstructed field has sharp interfaces between grains and grain boundaries, in contrast to the diffuse interfaces characteristic of the phase-field representation, in which $\eta_i$ varies smoothly across a finite-width boundary. We do not apply any smoothing to the reconstructed field prior to phase-field integration. Instead, the sharp reconstructed configuration is used directly as the initial condition, and the phase-field solver self-relaxes the interfaces to their equilibrium diffuse width during the first steps of the corrective simulation. Because the gradient-energy term in the free-energy functional (Eq. (\ref{eq:graingrowth})) penalizes sharp gradients in $\eta_i$, the interfaces rapidly evolve toward their equilibrium diffuse profile, restoring physical consistency before subsequent VFN prediction resumes. For spinodal decomposition, reconstruction is simpler, as the concentration field is obtained by interpolating the RGB values along the colormap to the interval $(0,1)$. This forward--reverse mapping establishes a bijective correspondence between phase-field data and image representation, enabling seamless transitions between simulation and prediction and ensuring that predicted images can be analyzed in the same manner as PF simulation outputs.

\subsection{Voxel flow blocks}
\label{sec:VFB}

The VFB is designed to refine voxel flow estimates in an end-to-end manner while avoiding restrictive assumptions such as locally linear or temporally symmetric motion. Given two consecutive input phase field snapshot frames $I_{t-1}$ and $I_t$, a synthesized intermediate prediction frame $\tilde{I}_{t+1}^{i-1}$, and the voxel flow predicted by the previous block $F^{i-1}_{t+1}$, the $i$-th VFB learns to approximate the target voxel flow $F^i_{t+1}$. The initial block starts from $\tilde{I}^{0}_{t+1}=0$ and $F^{0}_{t+1}=0$, and a sequence of nine VFBs is used to produce the final future frame. The detail of VFB is described in algorithm \ref{alg:VFB}.

To effectively capture both large-scale motion and fine spatial structures, each VFB uses a two-branch architecture inspired by pyramidal optical-flow estimation. The motion branch operates on a downsampled version of the input PF snapshot frames, expanding the receptive field for large displacements. For the $i$-th block, a scaling factor $S_i\in\{4,4,4,2,2,2,1,1,1\}$ is used, and both the input PF frames and the incoming voxel flow are rescaled by $1/S_i$. After feature extraction via two strided convolutions and residual convolutional blocks, the resulting motion features are compressed using a squeeze convolution and subsequently upsampled by a factor of $2S_i$.

In parallel, a spatial branch processes the original-resolution inputs using a light convolutional stack, preserving local textures and structural cues. The features from the two branches are concatenated and decoded using a transposed convolution to produce a 5-channel output comprising a refined voxel flow and a soft blending mask. This design allows each block to progressively refine motion estimates while maintaining spatial fidelity.

Across all nine VFBs, the receptive field grows naturally due to multiscale downsampling, while the spatial branch ensures that high-frequency details are retained. This coarse-to-fine refinement strategy enables the model to handle large, nonlinear, or asymmetric motions without the need for explicit flow linearization, symmetry assumptions, or flow-reversal layers.
\begin{algorithm}
\label{alg:VFB}
\caption{Multiscale Voxel Flow Block (VFB)}
\KwIn{Input frames $I_{t-1}, I_t$, previous synthesized frame $\tilde{I}^{i-1}_{t+1}$, previous voxel flow $F^{i-1}_{t+1}$, scale factor $S_i$}
\KwOut{Refined voxel flow $F^{i}_{t+1}$, blending mask $M^{i}$}

\BlankLine
\textbf{1. Input preparation:}\\
Concatenate inputs: $X \leftarrow [I_{t-1},\, I_t,\, \tilde{I}^{i-1}_{t+1}]$,  $F \leftarrow F^{i-1}_{t+1}$\;
\If{$S_i \neq 1$}{
    $X_d \leftarrow \mathrm{Interp}(X, 1/S_i)$\;
    $F_d \leftarrow \mathrm{Interp}(F, 1/S_i) / S_i$\;
}
\Else{
    $X_d \leftarrow X$, \quad $F_d \leftarrow F$\;
}

\BlankLine
\textbf{2. Motion branch (low resolution):}\\
$H_1 \leftarrow \mathrm{Conv}_{\downarrow}([X_d, F_d])$ \tcp*{two strided conv layers}
$H_2 \leftarrow \mathrm{ConvBlock}(H_1) + H_1$ \tcp*{three residual conv blocks}
$H_s \leftarrow \mathrm{Conv}_{sq}(H_2)$  \tcp*{channel squeeze}
$H_m \leftarrow \mathrm{Interp}(H_s,\, 2 S_i)$  \tcp*{upsample to final scale}

\BlankLine
\textbf{3. Spatial branch (full resolution):}\\
$H_{sp} \leftarrow \mathrm{Conv}([X,F])$\;
$H_{sp} \leftarrow \mathrm{ConvBlock}(H_{sp})$\;

\BlankLine
\textbf{4. Fusion and decoding:}\\
$H \leftarrow [H_m,\, H_{sp}]$ \tcp*{feature fusion}
$T \leftarrow \mathrm{ConvTranspose}(H)$ \tcp*{decode voxel flow + mask}

\BlankLine
\textbf{5. Output splitting:}\\
$F^{i}_{t+1} \leftarrow T[:, 1{:}4]$ \tcp*{4-channel voxel flow}
$M^{i} \leftarrow T[:, 5]$ \tcp*{1-channel blending mask}

\Return{$F^{i}_{t+1}, M^{i}$}
\end{algorithm}

\subsection{Routing mechanism and VFN architecture}
\label{sec:VFN}

The VFN predicts future PF evolution frames by progressively refining a voxel flow field through a sequence of nine VFBs. The output of each VFB updates the current voxel flow estimate, which is then used to warp the input frames and synthesize an intermediate prediction.

To decide which scales of VFBs to go through during one inference, VFN incorporates a dynamic routing mechanism that adaptively selects a sub-network for each input pair. Motion complexity varies across different frame pairs, and full-depth computation is unnecessary for small-motion samples. VFN therefore introduces a lightweight routing module to predict a binary routing vector indicating which VFBs should be executed. This design allows up to $2^9$ possible inference paths, enabling flexible computation tailored to input motion magnitude.

The routing module consists of two convolutional layers followed by global average pooling and a fully connected layer, producing a nine-dimensional probability vector $\tilde{v}$. The binary routing vector $v\in\{0,1\}^9$ that determines which VFBs contribute updates to the voxel flow is obtained by Bernoulli sampling of the soft routing weights through a straight-through estimator, so a block with a high weight is activated with high but not certain probability. During training, both activated and deactivated branches are averaged to maintain differentiability, while at inference only the activated blocks are executed. This dynamic strategy preserves representational capacity for large motions while significantly reducing computational cost for simpler inputs. The final synthesized frame is formed by warping the input frames with the refined voxel flow and blending them using the learned mask.
\begin{algorithm}
\label{alg:VFN}
\caption{Dynamic Routing and Forward Pass of VFN}
\KwIn{Input frames $I_{t-1}, I_t$, scaling factors $\{S_i\}_{i=1}^9$}
\KwOut{Synthesized future frame $\tilde{I}_{t+1}$}

\BlankLine
\textbf{1. Routing prediction:}\\
$F \leftarrow 0$, $M \leftarrow 0$, $\tilde{I} \leftarrow 0$ \tcp*{initial voxel flow, mask, and frame}
$R \leftarrow \mathrm{RoutingModule}(I_{t-1}, I_t)$ \tcp*{two convs + avg pooling}
$\tilde{v} \leftarrow \sigma(\mathrm{Linear}(R))$ \tcp*{routing probabilities}
$v \leftarrow \mathrm{RoundSTE}(\tilde{v})$ \tcp*{binary routing vector by Bernoulli sampling}

\BlankLine
\textbf{2. Iterative refinement through nine MVFB blocks:}\\
\For{$i = 1$ \KwTo $9$}{
    \If{\textbf{training}}{
        $\Delta F_i, \Delta M_i \leftarrow \mathrm{MVFB}_i([I_{t-1}, I_t, \tilde{I}, M], F;\ S_i)$\;
        $F_\text{temp} \leftarrow F + \Delta F_i$, \quad $M_\text{temp} \leftarrow M + \Delta M_i$\;
        $F \leftarrow F + v_i \cdot \Delta F_i$ \tcp*{weighted update}
        $M \leftarrow M + v_i \cdot \Delta M_i$\;
        Compute warped frames using both $F_\text{temp}$ and $F$\;
    }
    \Else{
        \If{$v_i = 1$}{
            $\Delta F_i, \Delta M_i \leftarrow \mathrm{MVFB}_i([I_{t-1}, I_t, \tilde{I}, M], F;\ S_i)$\;
            $F \leftarrow F + \Delta F_i$, \quad $M \leftarrow M + \Delta M_i$\;
        }
    }
    Warp $I_{t-1}$ and $I_t$ using the updated $F$ to obtain $\tilde{I}$\;
}

\BlankLine
\textbf{3. Final synthesis:}\\
$\tilde{I}_{t+1} \leftarrow \tilde{I} \cdot \sigma(M) + (1 - \sigma(M)) \cdot$ opposing warp\;
\Return{$\tilde{I}_{t+1}$}
\end{algorithm}

\subsection{Phase field simulation}
\label{sec:PF}

To test the generalizability of our framework, we selected two phase-field models to build the training, validation, and test datasets: the grain-growth model of L. Q. Chen \cite{chen1994computer} and the spinodal decomposition model of A. Cahn \cite{cahn1961spinodal}. For the grain-growth model, the initial microstructure was generated using Voronoi tessellations, and a regularization method \cite{regularization} with regularity parameter $\alpha=0.8$ was used to ensure that the grains generated by the Voronoi tessellation followed a more realistic area distribution. The image size was 512$\times$512, and the number of grains in each image was 112. The grain-growth phase-field model is formulated as follows: a total free-energy functional composed of bulk and interfacial free-energy density integrals, as shown in equation (\ref{eq:graingrowth}).
\begin{equation}
    F(\eta_i(\boldsymbol{r})) = \int_V \left[ \sum_{i=1}^{N_g}\left( -\frac{\eta_i^2}{2}+\frac{\eta_i^4}{4} \right)+\sum_{i=1,j>i}^{N_g}\eta_i^2\eta_j^2 + \sum_{i=1}^{N_g}\frac{\kappa}{2}|\nabla\eta_i|^2 \right]\mathrm{d}V
    \label{eq:graingrowth}
\end{equation}
Where $\eta_i$ are the order parameters of the model, representing grain $i$ existence when $\eta_i=1$, and nonexistence when $\eta_i=0$. The gradient energy coefficient is $\kappa=1$. The evolution for this model was defined with the Allen-Cahn equation:
\begin{equation}
    \frac{\partial\eta_i}{\partial t} = -L_i\frac{\delta F}{\delta\eta_i}
    \label{eq:allencahn}
\end{equation}
where $L_i$ are kinetic coefficients, here we use isotropic configuration to let all $L_i$ equal to each other. The model parameters of the grain growth snapshots were $L=1$, $\Delta t=0.2$, $\Delta x=2$, $\kappa=2$, $\alpha=1$, $\beta=1$, $\gamma=1$. \cite{chen1994computer} The snapshots were visualized using the turbo colormap, in which colors were assigned according to the phase-field order-parameter index of each grain, following the color-phase-field data mapping strategy described in section \ref{sec:framework}. 

For the spinodal decomposition model, random initial configurations were employed. We used regular solution model as the Helmholtz free energy term, added with gradient energy, the free energy density writes:
\begin{equation}
    f(c)=\ln{c}+\ln{(1-c)}+\Omega c(1-c) + \kappa(\nabla c)^2
\end{equation}
where $\Omega=3.4, \kappa=1$, for evolution, diffusivity $D=0.05$ and $\Delta t=0.1$. Periodic boundary conditions were applied in both simulations. The two PF models were implemented with explicit finite difference method, and paralleled with NVIDIA numba-cuda package. The PF simulations in the framework are all performed with GPU computation.

\subsection{Dataset preparation and model training}

To train, validate, and test the VFN within the prediction framework, we constructed datasets from evolution snapshots generated by these two PF models. Each dataset consists of multiple simulations of a single PF model, with each simulation initialized from a different random condition, as indicated by $\times n$ in figure \ref{fig:method}(e). For every simulation, snapshots were then recorded sequentially at uniform intervals of 1,000 PF time steps along the temporal direction $t$.

The framework was implemented and trained using PyTorch \cite{paszke2019pytorch}, and all numerical operations and data handling were performed using NumPy \cite{harris2020array}. The model was trained separately for the grain-growth and spinodal-decomposition phase-field models. For grain growth, the training and validation datasets were used during model optimization and training monitoring. An independent test dataset containing 20 PF simulations, each with 20 images, was reserved for the quantitative evaluations and representative results reported in the Results section. We trained the model for 1,500 epochs on an NVIDIA RTX A4000 GPU. The training batch size was 28, and cosine annealing was used to reduce the learning rate from $10^{-4}$ to $10^{-5}$. The training loss was defined as the sum of the $l_1$ loss and the VGG loss. The $l_1$ loss was computed on the Laplacian pyramid representations, and the total loss function was their weighted sum, with parameter $\gamma=0.8$:
\begin{equation}
    L_{total} = \sum_{i=1}^n \gamma^{n-i}l_1(\tilde{I}_{t+1}^i, I_{t+1}) + 0.5l_{VGG}(\tilde{I}_{t+1}^9, I_{t+1})
\end{equation}
The training loss plot (figure S1 and S2) of VFN on grain growth and spinodal decomposition datasets are available in supplementary information.

\begin{figure}[pos=htbp]
    \centering
    \includegraphics[width=0.9\linewidth]{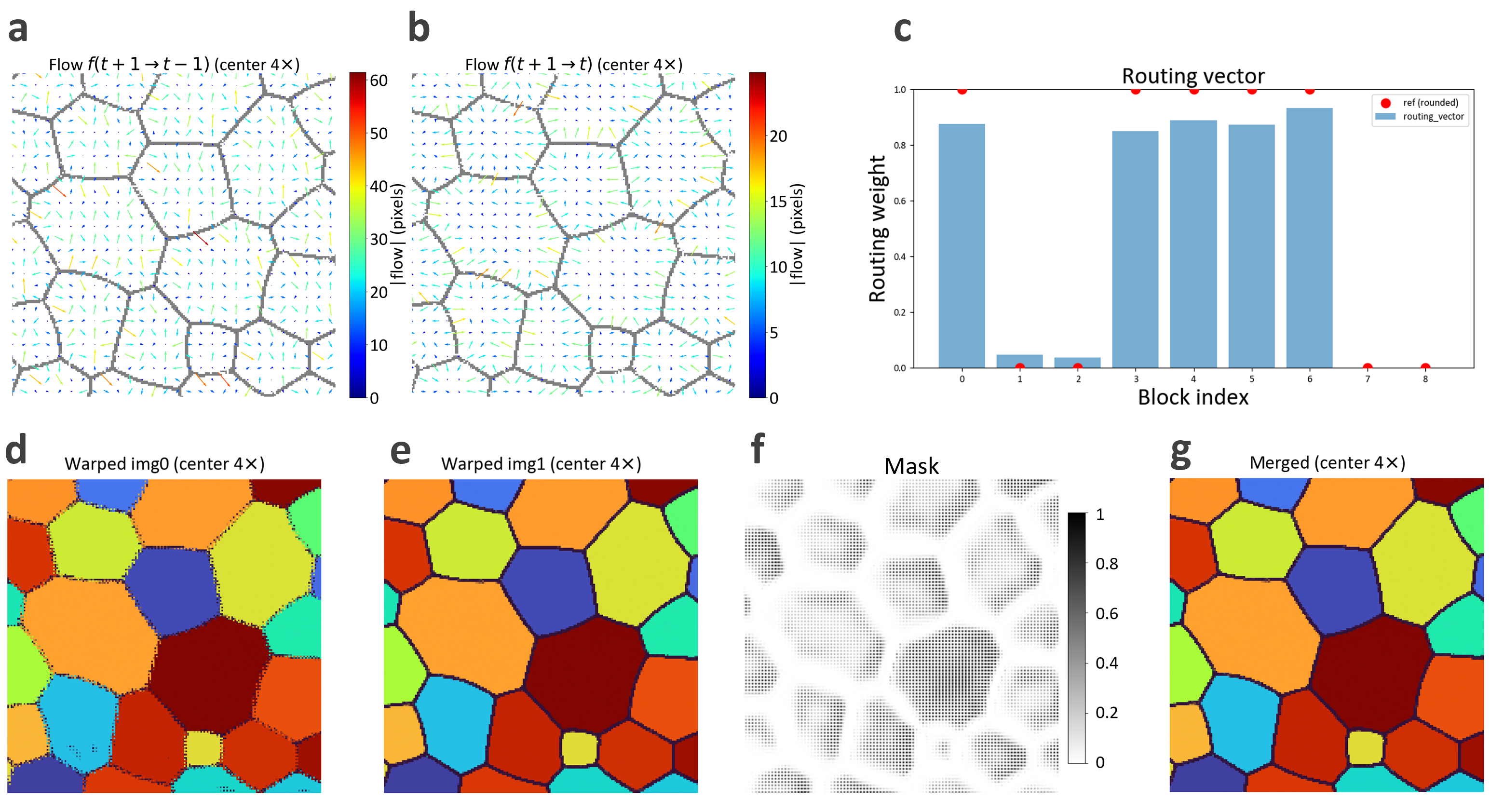}
    \caption{Internal prediction pipeline for a representative grain-growth step. (a,b) the activated blocks estimate backward flows $f_{t+1\to t-1}$ and $f_{t+1\to t}$ (arrows: direction; color: magnitude); (c) Routing vector selects which multiscale blocks run; (d,e) the warped image by these flows from the two input frames; (f) blending mask $m$; (g) merged prediction $I_{t+1}=m \times \overleftarrow{W}(I_{t-1}, f_{t+1 \to t-1})+(1 - m) \times \overleftarrow{W}(I_t, f_{t+1 \to t})$.}
    \label{fig:flow}
\end{figure}

To elucidate the internal mechanism of the VFN during microstructure-evolution prediction, we visualize the routing vector, the learned flow fields, the warped frames, the blending mask, and the final merged prediction for a representative grain-growth step in figure \ref{fig:flow}. Together these panels trace the full prediction pipeline: the routing vector first selects which multiscale blocks are activated; the activated blocks estimate the bidirectional flow fields; the two input frames are warped by their respective flows; and the warped frames are combined through the learned mask to produce the predicted frame. The prediction begins with the routing module, which produces the routing vector shown in figure \ref{fig:flow}(c). The routing module outputs a single nine-dimensional vector per frame that determines which of the nine voxel-flow blocks (VFBs) are executed. For the step shown, a subset of the blocks (blocks 0, 3, 4, 5 and 6) is activated, while the remaining blocks stay inactive. The temporal evolution of the routing vector over a long prediction trajectory, and its relationship to the coarsening microstructure, is analyzed separately in Section \ref{sec:routing_time}. The activated blocks estimate two backward flow fields, shown in figure \ref{fig:flow}(a) and (b). Each flow maps pixels of the predicted frame back to one of the two input frames; the arrows encode flow direction and the color encodes flow magnitude. The physically meaningful flow is concentrated along the grain boundaries, where the arrows align with the local boundary-normal direction. This pattern is consistent with the curvature-driven motion of grain boundaries: in the present isotropic grain-growth model, boundary migration is governed by local curvature, with the strongest displacements occurring along the most strongly curved, rapidly migrating boundary segments. We note that $f_{t+1\to t-1}$ spans two snapshot intervals while $f_{t+1\to t}$ spans one, so the former exhibits correspondingly larger magnitudes for the same underlying motion. The two input frames are then warped by their respective flow fields, producing the warped frames in figure \ref{fig:flow}(d) and (e). These warped frames are combined into the final prediction through the learned blending mask $m$ shown in figure \ref{fig:flow}(f). The mask is a per-pixel weight in $[0,1]$ that determines, at each location, how much of the prediction is drawn from the frame warped from $I_{t-1}$ versus the frame warped from $I_t$; the predicted frame is formed as $I_{t+1}=m \times \overleftarrow{W}(I_{t-1}, f_{t+1 \to t-1})+(1 - m) \times \overleftarrow{W}(I_t, f_{t+1 \to t})$. The resulting merged prediction is shown in figure \ref{fig:flow}(g), which reproduces the grain morphology of the target frame, confirming that the warp-and-blend mechanism yields a physically consistent prediction.

\section{Results}

\subsection{Voxel flow network prediction of grain growth}

\begin{figure}[pos=htbp]
    \centering
    \includegraphics[width=0.8\linewidth]{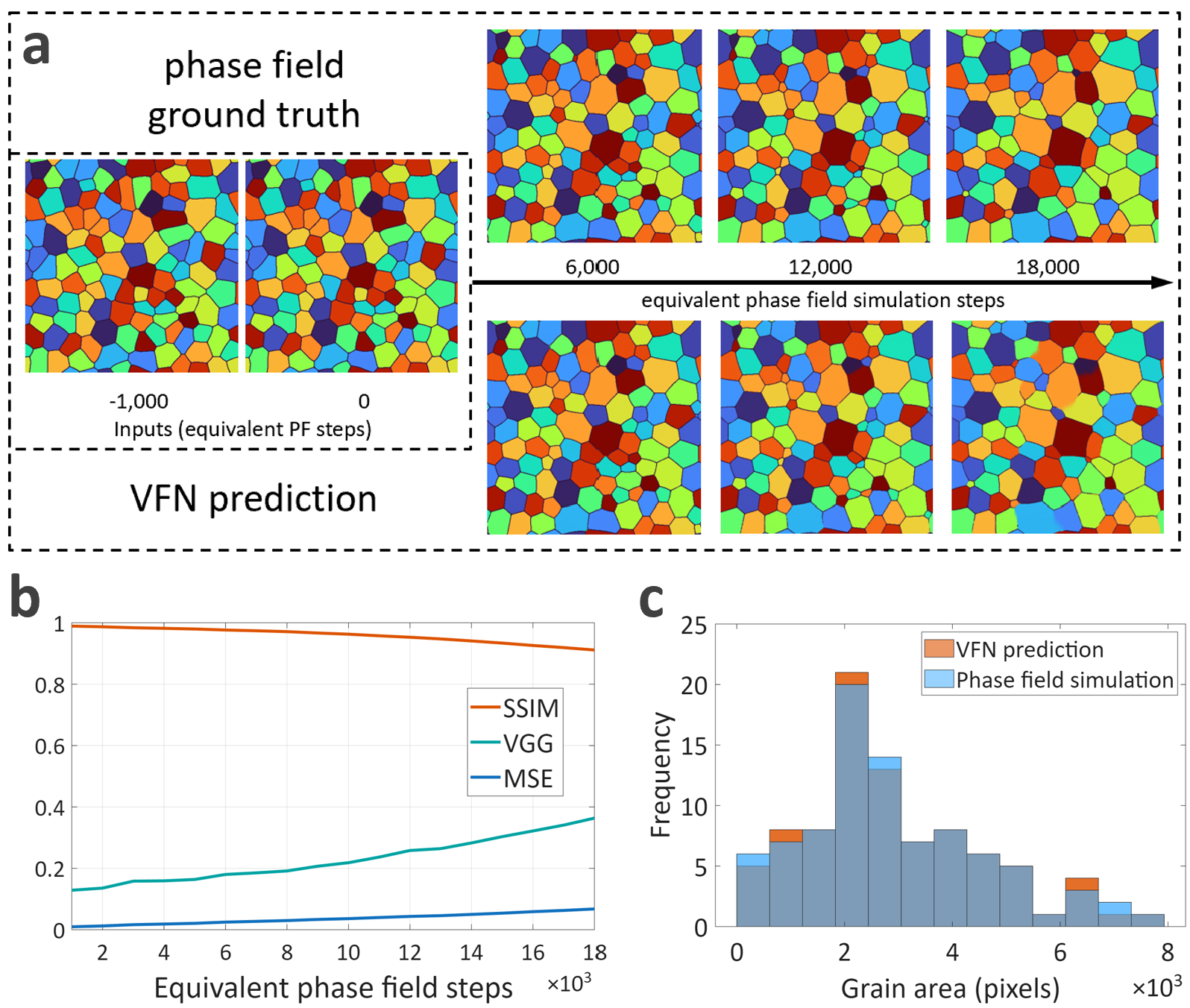}
    \caption{Voxel-flow network prediction results. (a) Input images, ground-truth images, and predicted images of grain-growth evolution (the full 18-frame sequence is shown in figure S3); (b) SSIM, MSE, and VGG loss curves; and (c) grain-area distributions of the final image from VFN prediction and phase-field simulation.}
    \label{fig:vfngrain}
\end{figure}

We applied the VFN to the grain-growth phase-field model (see section \ref{sec:PF} for the formulation and parameters), and the test results are shown in figure \ref{fig:vfngrain}. The resolution of predicted snapshots was 512$\times$512, and the interval between successive phase-field snapshots was 1,000 time steps. For inputs, the model was provided with two simulation images separated by 1,000 time steps. The initial configuration contained 112 grains, shown in the left rectangle in figure \ref{fig:vfngrain}(a). The network then predicted 18 consecutive images, corresponding to 18,000 phase-field time steps. After these 18 predicted images, the grain number decreased from 112 to 88. Three representative images from both the prediction and the PF simulation are shown on the right side of figure \ref{fig:vfngrain}(a). The predicted results are displayed on the top row and aligned vertically with their corresponding ground-truth simulations on the bottom row. This comparison demonstrates that the voxel-flow network achieves high accuracy in predicting grain-growth dynamics on the test dataset. 

\subsubsection{Quantitative accuracy metrics}

To further quantify the accuracy of the network, we evaluated the predictions using three metrics: mean squared error (MSE), VGG perceptual loss \cite{perceptualloss}, and structural similarity index (SSIM) \cite{wang2004image}. The MSE was NOT computed in image RGB space, but computed in the extracted phase field configurations pointwise as follows. Since each point in the simulation domain belongs either to a grain interior or to a grain boundary, we defined the local error using a delta function and then aggregated it across the domain. This would give more physical information than using image MSE to reflect the prediction quality.
\begin{equation}
    e(x,y) = \delta_{ij}(x,y), \quad
\begin{cases}
i = \text{phase index in predicted image at } (x,y), \\
j = \text{phase index in PF simulation at } (x,y),
\end{cases}
\end{equation}
\begin{equation}
    \mathrm{MSE} = \frac{1}{W \times H} \sum_{x=1}^{W} \sum_{y=1}^{H} e(x,y)^2
\end{equation}
For the perceptual loss, we employed the pretrained VGG19 network. SSIM values were computed using the scikit-image package \cite{scikit-image}. The averaged results over the test dataset containing 20 simulations are shown in figure \ref{fig:vfngrain}(b). As expected, MSE increased approximately linearly with prediction length, while VGG loss increased and SSIM degraded more rapidly. This behavior likely reflects error accumulation in the iterative prediction scheme, where predicted frames are repeatedly used as inputs for subsequent steps. After 18 predicted frames, the metrics still maintained 6.72\% (MSE), 0.360 (VGG loss), and 0.911 (SSIM). 

We also compared macroscopic statistics between the predicted configurations and their phase-field ground truths. Specifically, phase-field data were reconstructed from predicted images, and the grain area distribution was computed. A histogram of grain areas, shown in figure \ref{fig:vfngrain}(c), compares the 18th predicted image with the corresponding simulation result. Among the 88 grains present, the distributions differed by only a magnitude of one at three histogram bins, while the total bin number was 13. This consistency demonstrates that the predicted images accurately reproduce the size and morphology of individual grains, further validating the network's performance.

\subsubsection{Prediction results on topological features}

\begin{figure}[pos=htbp]
    \centering
    \includegraphics[width=0.8\linewidth]{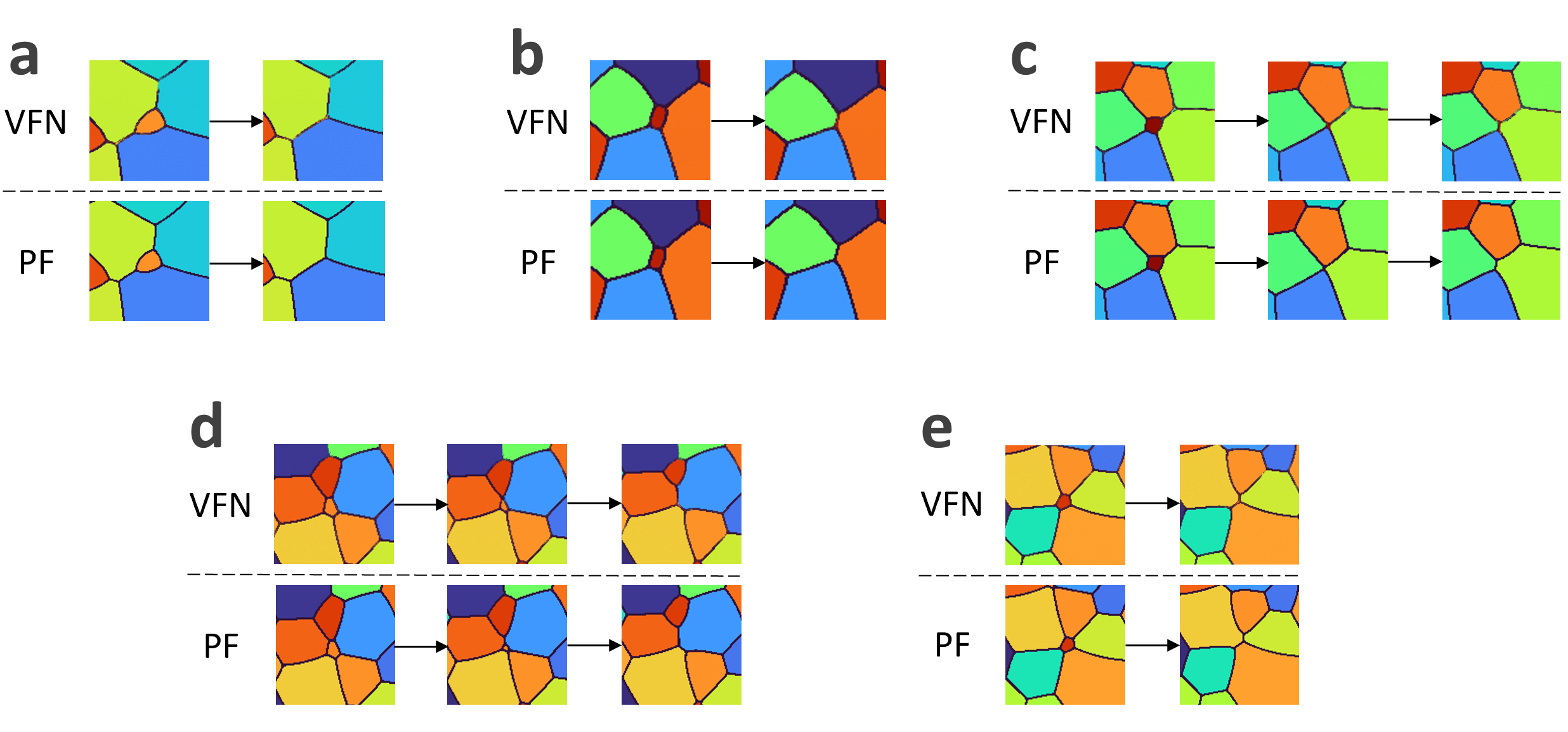}
    \caption{VFN predictions of grain-disappearance events, compared with PF ground truth. (a) Disappearance of a 3-sided grain. (b) Disappearance of a 4-sided grain with one elongated direction. (c) Disappearance of a nearly square 4-sided grain, producing a 4-grain junction. (d) A 5-sided grain first transforms into a 4-sided grain before disappearing. (e) Disappearance of a 5-sided grain, producing a 5-grain junction.}
    \label{fig:vanish}
\end{figure}

Over this relatively long timescale, the predictions correctly captured key topological features such as grain disappearance, grain-boundary migration, preservation of grain identity, and preservation of periodic boundaries. For grains that disappeared during prediction and simulation, the detailed grain-boundary morphology was also well preserved. In this prediction/simulation sequence, 24 grains vanished. We examined the snapshots at the moments of disappearance and categorized the events according to the number of neighboring grains surrounding the vanishing grain. A 3-sided grain represents the smallest topological configuration observed immediately before disappearance as shown in figure \ref{fig:vanish}(a). For 4-sided grains, two 3-grain junctions typically form at the location where the grain disappears, making it critical for the model to predict the correct orientation of these junctions. In some cases, one dimension of the 4-sided grain is much longer than the perpendicular one before disappearance; in such cases, the longer direction becomes a grain boundary, with two junctions forming at its ends, as illustrated in figure \ref{fig:vanish}(b). Other 4-sided grains remain nearly square before vanishing, producing a 4-grain junction at the disappearance site; the model must then predict the direction in which this 4-grain junction splits into two 3-grain junctions, as shown in figure \ref{fig:vanish}(c). For grains with five or more sides, the process is similar to that of 4-sided grains. They either first reduce to 4-sided grains before vanishing (figure \ref{fig:vanish}(d)), or remain 5-sided until disappearance, generating a 5-grain junction that subsequently transforms into three 3-grain junctions connected by two grain-boundary segments (figure \ref{fig:vanish}(e)). In the cases discussed above, the figures show that the VFN predictions reproduce the correct topological class of these disappearance events, including the orientation in which higher-order junctions resolve into three-grain junctions. However, the VFN does not always predict such topological features with perfect accuracy. In the 18-frame prediction shown in figure \ref{fig:vfngrain}, 24 grains vanished over the course of the evolution. Of these 24 disappearance events, 22 were topologically correct, with 18 occurring at the correct time and a further 4 reproduced correctly but with a timing offset (3 later and 1 earlier than the ground truth); only 2 were genuine topological errors, corresponding to spurious disappearances caused by loss of grain-boundary definition. The full sequence is provided in supplementary figure S3. These timing and topological errors, although affecting only a minority of events, accumulate over long prediction horizons and underscore the need for an error-control mechanism such as the periodic phase-field correction which we will discuss later.

Compared with similar efforts in previous studies, our results are more robust because grain identity is explicitly tracked and preserved through color mapping, thereby retaining the complete physical configuration of the grain system. Although minor defects were observed—for instance, distortions in grain shape after long prediction horizons and partial grain boundary smearing—the overall performance of the network is satisfactory. Moreover, although conceptually simple, the padding method effectively preserved the periodic boundary features in the input images, reflecting the effectiveness of the boundary treatment. We speculate that this improvement arises because voxel flows at boundaries are inherently more difficult to capture than those in interior regions.

\subsubsection{Comparison with other methods}

We conducted a systematic comparison between the proposed VFN architecture and several representative spatiotemporal prediction methods, including ConvLSTM \cite{RN100} and SimVP \cite{RN99}, to evaluate their capability for long-term microstructure evolution prediction. A primary advantage of the VFN lies in its lightweight model design. When trained on the same grain growth dataset, the competing methods exhibit significantly larger model sizes and higher memory demands than VFN, which restricts their practical applicability, particularly for large-scale simulations or resource-constrained computing environments. In contrast, the compact architecture of VFN enables efficient deployment without sacrificing prediction accuracy. A second key advantage of VFN is its computational efficiency in both training and inference. During training, the time required per epoch for VFN is substantially shorter than that of ConvLSTM and SimVP, reflecting the efficiency of its voxel-flow-based formulation. During inference, benchmark tests \cite{tan2023openstl} further show that VFN achieves orders-of-magnitude higher frame rates than the other methods, while also requiring fewer floating-point operations (FLOPs). This high inference throughput is particularly important for long-horizon prediction tasks, where computational cost can otherwise become prohibitive.

\begin{figure}[pos=htbp]
    \centering
    \includegraphics[width=.75\linewidth]{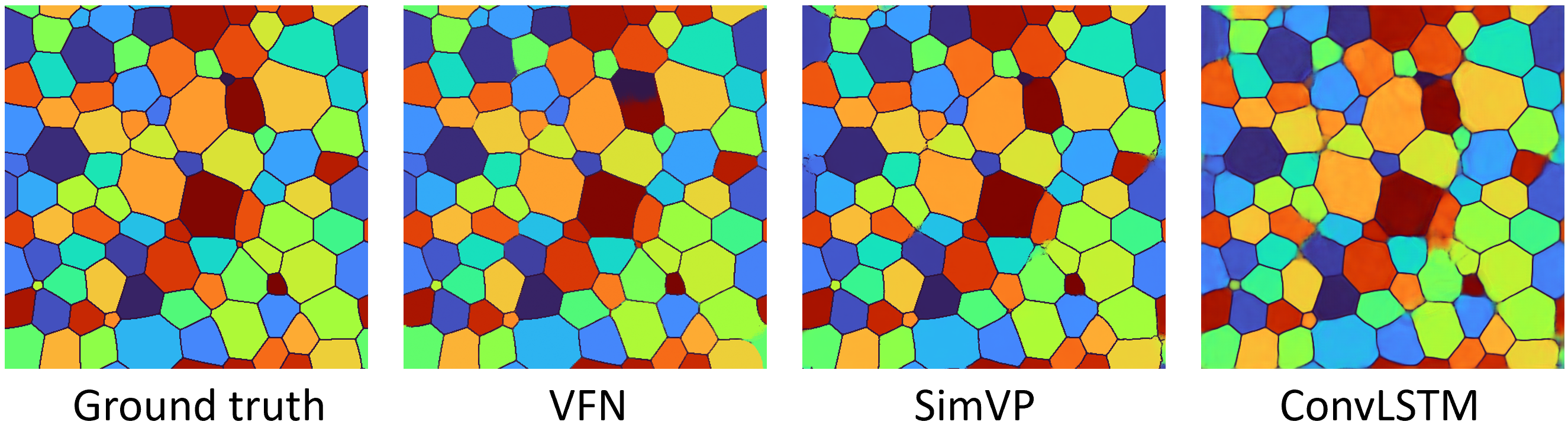}
    \caption{Comparison of the 20th snapshot produced by different prediction methods. VFN was configured with a 2-18 input-output setting, whereas the other methods used a 10-10 input-output setting.}
    \label{fig:comparison}
\end{figure}

To assess predictive performance, we compared the MSE, SSIM and VGG perceptual loss of grain growth predictions generated by the different models using the OpenSTL \cite{tan2023openstl} platform. Due to differences in the default model configurations, we adopted distinct input-output settings for ConvLSTM and SimVP. Specifically, these models were trained using 10 input frames to predict the subsequent 10 frames, rather than the more challenging 2-to-18 frame prediction setting used by the VFN. The predicted results for one representative simulation, together with the corresponding ground truth, are shown in figure \ref{fig:comparison}; this case is one of the 20 simulations in the test set, and is shown for illustration only. Despite operating under a significantly more demanding prediction setting, the VFN achieves superior performance. As summarized in table \ref{tab:comparison}, which reports the mean and standard deviation of each metric computed from the final predicted (20th) image across all 20 test simulations, the VFN consistently outperforms the other models across the evaluated metrics. The low standard deviation of the VFN metrics across the 20 test simulations (e.g., MSE of $6.72 \pm 1.19\%$) indicates consistent performance across different initial microstructures, whereas the baselines exhibit both higher mean error and larger variability (SimVP MSE $32.88 \pm 10.23\%$). Although not immediately apparent in figure \ref{fig:comparison}, the SimVP and ConvLSTM predictions do not consistently preserve the original grain colors, which leads to larger error metrics than might be expected.

\begin{table}
    \centering
    \begin{tabular}{cccc}
    \hline
        Method & VFN & SimVP & ConvLSTM \\
    \hline
        MSE  & $6.72 \pm 1.19\%$   & $32.88 \pm 10.23\%$ & $30.51 \pm 4.73\%$ \\
        SSIM & $0.911 \pm 0.016$   & $0.902 \pm 0.012$   & $0.882 \pm 0.013$ \\
        VGG  & $0.360 \pm 0.044$   & $0.568 \pm 0.050$   & $0.583 \pm 0.045$ \\
    \hline
    \end{tabular}
    \caption{Comparison of error metrics on the 20th (final predicted) image, reported as mean $\pm$ standard deviation over the 20 simulations in the test set. Note that VFN was evaluated under a harder 2-to-18 input-output setting, whereas SimVP and ConvLSTM used a 10-to-10 input-output setting; the comparison is therefore not strictly like-for-like and favors the baselines.}
    \label{tab:comparison}
\end{table}

\subsection{Joint prediction results}

From the VFN prediction results, we observe that while the model achieves high accuracy for short-term evolution, errors accumulate approximately linearly with the number of predicted images. More critically, non-physical artifacts emerge, such as grain boundary dissolution at longer horizons. For example, in figure \ref{fig:joint}(a), the 38th VFN-predicted image exhibits blurred and irregular boundaries that deviate from physical expectations. To extend the predictive horizon while maintaining physical fidelity, we adopt the joint prediction scheme introduced in figure \ref{fig:method}(b).

\subsubsection{Joint prediction on 38 frames of grain growth}

\begin{figure}[pos=htbp]
    \centering
    \includegraphics[width=0.8\linewidth]{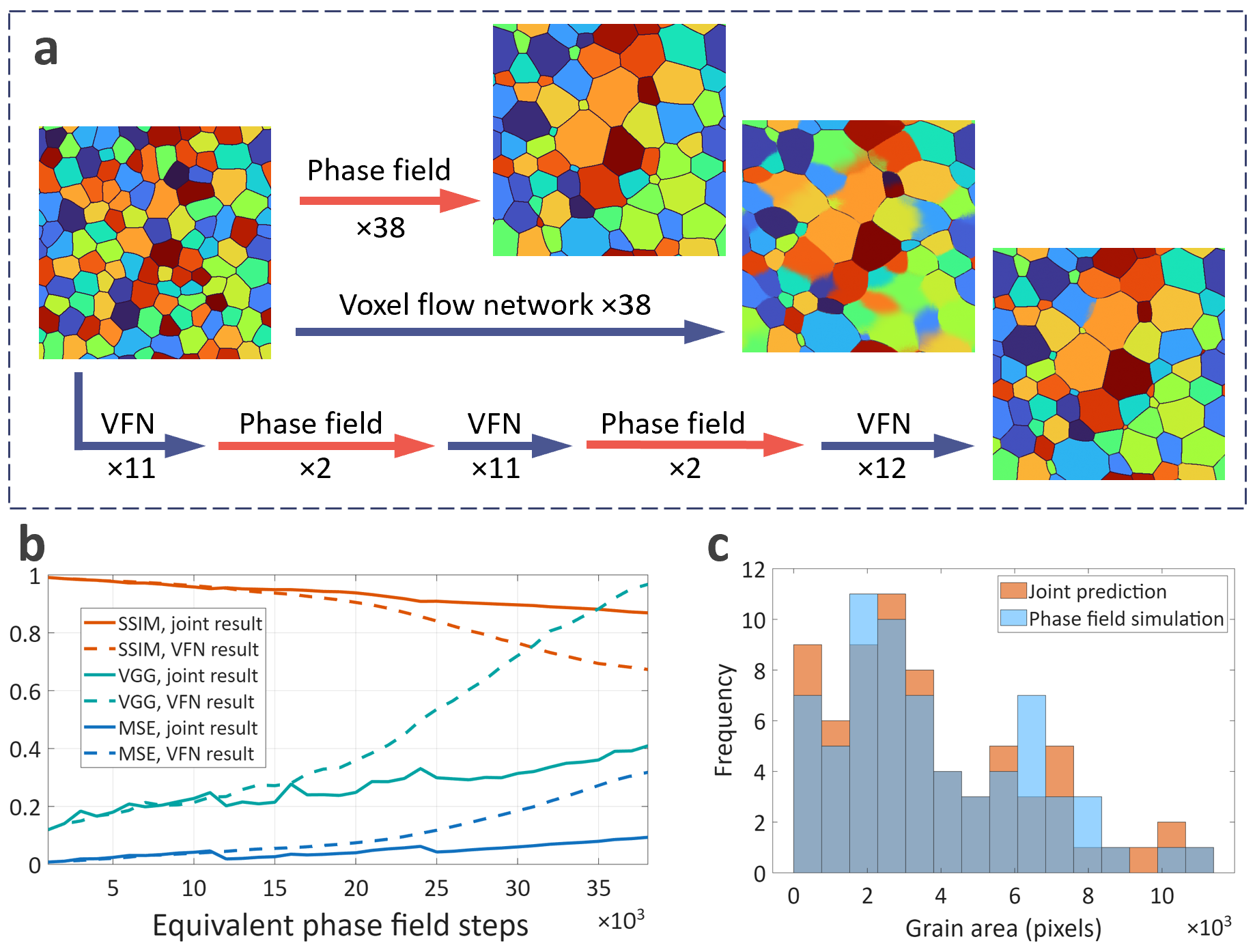}
    \caption{Joint prediction of grain growth. (a) Comparison of phase-field ground truth (red background), VFN-only prediction (blue background), and joint prediction (green background) over 38 frames. The 38th VFN-only prediction shows significant non-physical grain boundary dissolution, while the joint approach maintains grain shape and preserves physically realistic boundaries. (b) Quantitative evaluation of prediction accuracy over 38 frames using MSE, VGG perceptual loss, and SSIM. Dashed lines indicate VFN-only predictions, and solid lines indicate joint predictions. Sharp decreases in error correspond to inserted phase-field simulation segments, illustrating the corrective effect of the joint scheme. (c) Grain-area distributions of the final image from joint prediction and phase-field simulation.}
    \label{fig:joint}
\end{figure}

In this scheme, we alternate between VFN predictions and short bursts of phase-field simulation, thereby stabilizing grain morphology and reducing accumulated error. Specifically, for a 38-image sequence (equivalent to 38,000 PF steps), we interleave two PF segments of 2000 steps each, producing two frames required for the next VFN input. With this setup, the prediction sequence is divided into 11 VFN - 2 PF - 11 VFN - 2 PF - 12 VFN, totaling 38 frames. Figure \ref{fig:joint}(a) compares the PF ground truth, the VFN-only prediction, and the joint prediction over the same horizon. The joint approach clearly preserves grain shape while suppressing nonphysical dissolution.

The error metrics reinforce this observation. As shown in figure \ref{fig:joint}(b), the dashed lines (VFN-only) show steady error growth, while the solid lines (joint) exhibit sharp reductions at the 12th and 25th frames, coinciding with PF insertions. Compared with VFN-only predictions, joint prediction reduces MSE from 31.89\% to 9.32\%, VGG loss from 0.967 to 0.409, and improves SSIM from 0.673 to 0.869. The comparison of grain-area distribution histograms between the joint prediction and the PF ground truth (figure \ref{fig:joint}(c)) further confirms the statistical agreement of the predicted results. These results confirm that short PF corrections act as “error resets,” enabling stable long-term prediction without compromising efficiency. 

We also examined the effect of the length of each PF correction, that is, the number of phase-field integration steps inserted at each correction. Holding the correction frequency essentially fixed, we compared an 18-2-18 profile, in which each insertion runs 2,000 phase-field steps and returns two frames, with a 17-4-17 profile, in which each insertion runs 4,000 steps and returns four frames. On the final predicted frame, the longer 4,000-step insertion gave a modestly lower error (MSE 0.1073, SSIM 0.8542, VGG 0.4790) than the 2,000-step insertion (MSE 0.1176, SSIM 0.8383, VGG 0.4959). Longer phase-field steps therefore yield a small improvement in accuracy, consistent with the reconstructed sharp field relaxing more fully toward its equilibrium diffuse-interface width, but the effect is minor relative to the additional computational cost. Because the VFN requires two consecutive frames as input, a two-frame (2,000-step) insertion is the minimum that supplies a complete and physically consistent input pair for the subsequent prediction. We therefore adopt two-frame phase-field insertions as the default, as they provide a solid physical input to the VFN while keeping the correction cost low.

A natural question is whether the phase-field correction merely reconstructs the grain boundaries blurred by repeated VFN iterations, or whether the PF integration additionally improves the physical fidelity of subsequent predictions. Two observations indicate that the correction does more than sharpen boundaries. First, as shown in figure \ref{fig:joint}(b), each phase-field insertion not only produces a sharp local reduction in error but also lowers the rate at which error subsequently accumulates: the prediction error following an insertion grows more slowly than in the uncorrected rollout. A purely image-based reconstruction, such as boundary sharpening or denoising, would restore the appearance of the corrected frame but would leave the subsequent error-growth rate unchanged, since it does not alter the underlying physical state. Second, increasing the number of integration steps per insertion further reduces the error as discussed in previous paragraph, which would not occur if the correction acted only as a sharpening operation. The reason is that after many iterations VFN predictions drift away from the manifold of physically admissible microstructures, with smeared interfaces and degraded local equilibrium. A short phase-field integration relaxes the reconstructed field back onto the solution of the governing Allen-Cahn equations, restoring the equilibrium diffuse-interface profile, the curvature-driven boundary velocities, and the triple-junction configurations, so that the frames returned to the network are physically self-consistent. The phase-field correction therefore controls physical error rather than merely reconstructing blurred boundaries, which is a property that image post-processing alone cannot provide.

\subsubsection{Joint prediction results of grain growth over ultra-long time scales}
\label{sec:ultralong}

\begin{figure}[pos=htbp]
    \centering
    \includegraphics[width=0.8\linewidth]{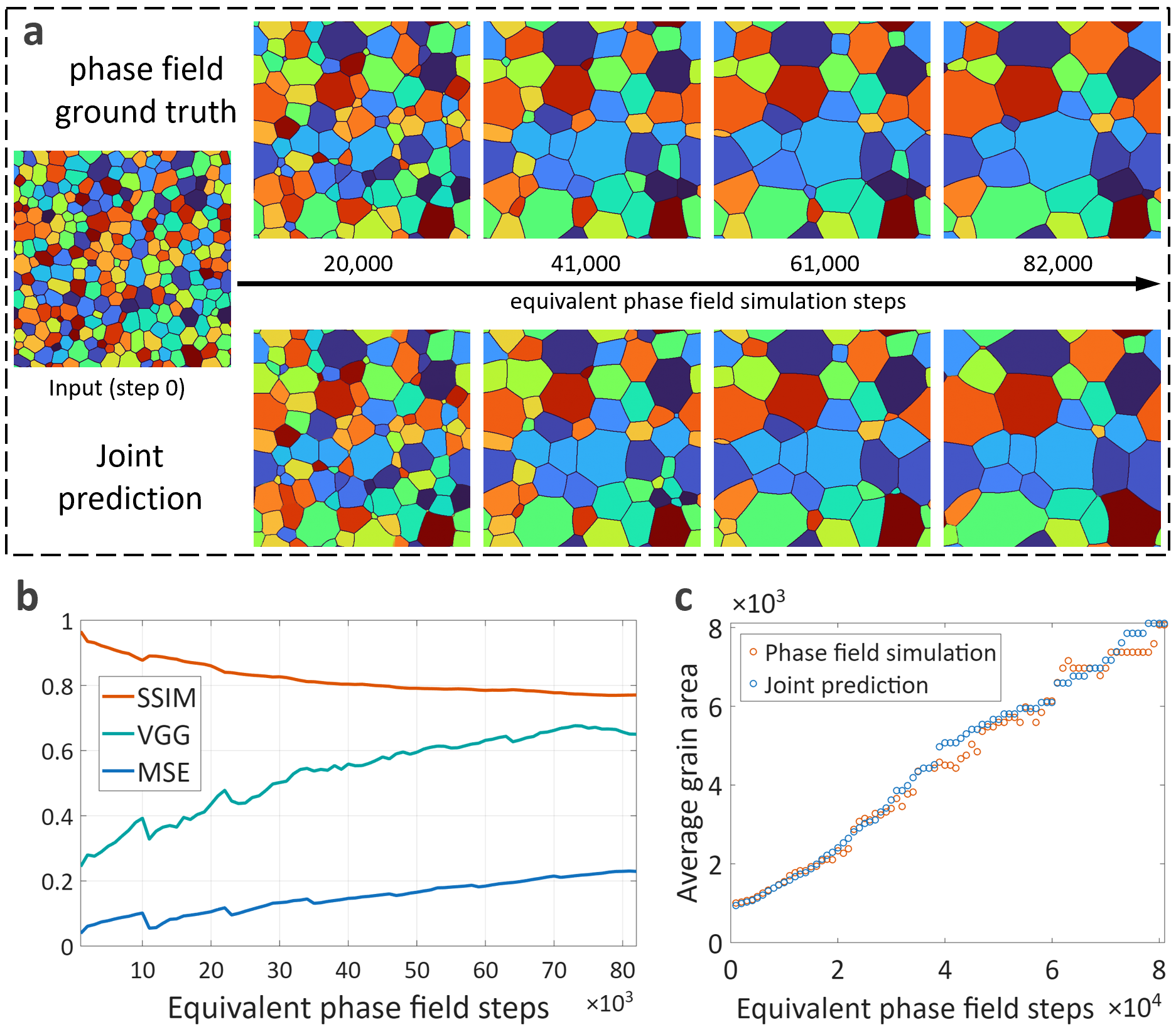}
    \caption{Long-term joint prediction for large grain microstructure. (a) Comparison of phase-field simulation (bottom) and joint-predicted (top) configurations for an initial microstructure containing 600 grains over 82 frames (full 82 frames in figure S5). Fine-grained details differ, but overall grain morphology and area distribution are well preserved. (b) Prediction error metrics (MSE, VGG loss, and SSIM) over 82 frames. Phase-field insertions reduce error locally, while the error growth rate decreases over time as grain coarsening slows the dynamics. (c) Statistical comparison of average grain area between phase-field simulation and joint prediction. Early evolution (frames 1-30) matches closely, while later evolution (frames 31-82) exhibits small oscillations due to discrete grain elimination events. Overall, the joint prediction captures the long-term trend and variability accurately.}
    \label{fig:grainlong}
\end{figure}

To further test the model's long-horizon capability and consistency with physical evolution, we perform joint prediction for extremely long time range, with a dense initial microstructure containing 600 grains. The joint prediction model was run for 82 frames (82,000 PF steps), as total grain number decreases from 600 to 29 (which represents a huge time range), with results shown in figure \ref{fig:grainlong}(a). At this scale, fine-grained details differ between prediction and ground truth, but overall grain morphology and area distribution remain consistent.

Figure \ref{fig:grainlong}(b) shows the error metrics across the sequence. While PF insertions continue to reduce error locally, the overall growth rate of error decreases as the system coarsens. This reflects the physical fact that as grains enlarge, the evolution becomes slower and less sensitive to local fluctuations. To quantify statistical consistency, we evaluate the average grain area over time (figure \ref{fig:grainlong}(c)). Let $y_i, \hat{y}_i$ be the predicted and ground truth of average grain area over time, and $\bar{y}$ the average of $\hat{y}$. We calculated the NMSE of predicted average grain area following equation (\ref{eq:nmsegrainlong}):
\begin{equation}
    \label{eq:nmsegrainlong}
    \mathrm{NMSE} = \frac{\sum_{i=1}^{n} (y_i - \hat{y}_i)^2}{\sum_{i=1}^{n} (y_i - \bar{y})^2} = 1.64\%
\end{equation}
This proves that the trend of grain growth is well captured. Observing more specifically on the plot, the average grain area grows from 437 to 8687 pixels across the evolution. For the first 30 frames, joint and PF results match almost perfectly. Beyond 30 frames, the two curves oscillate but follow the same overall growth trend. The oscillations arise from discrete grain elimination events, which produce plateau-jump behavior in grain area statistics. Importantly, the joint framework captures both the trend and variability, demonstrating its robustness for long-term predictions. 

\section{Discussion}

\subsection{Temporal evolution of the routing vector weights}
\label{sec:routing_time}

\begin{figure}[pos=htbp]
    \centering
    \includegraphics[width=0.7\linewidth]{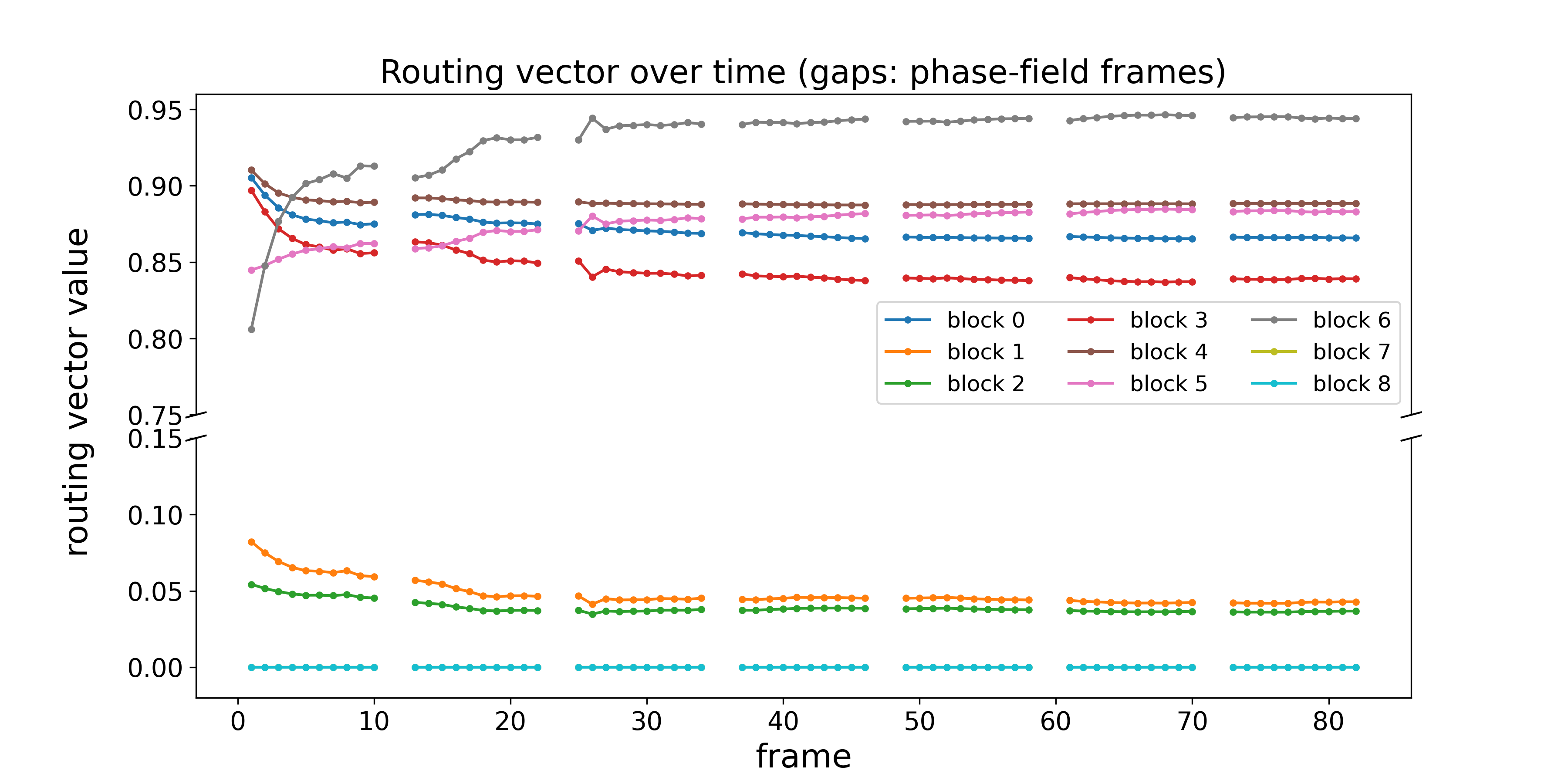}
    \caption{Soft routing-vector weights over a long joint prediction described in section \ref{sec:ultralong} (gaps: phase-field frames). Under scale list [4,4,4,2,2,2,1,1,1], blocks 0--2/3--5/6--8 are coarsest/intermediate/finest. During early rapid coarsening, finest-scale block 6 rises ($\sim$0.80$\to$0.94) and block 3 falls ($\sim$0.90$\to$0.84); weights then stabilize. Blocks 7--8 stay inactive.}
    \label{fig:routingtimeline}
\end{figure}

Figure \ref{fig:routingtimeline} shows the soft routing-vector weights over a long joint prediction described earlier in section \ref{sec:ultralong}, captured at every VFN-prediction step; the gaps along the time axis correspond to the inserted phase-field correction frames, during which the VFN is not invoked. Two regimes are apparent. During the early stage, while the microstructure is fine and coarsening is most rapid, several weights change appreciably: the weight of block 6 rises from approximately 0.80 to 0.94, while block 3 decreases from approximately 0.90 to 0.84, and the minor blocks (blocks 1 and 2) decay from roughly 0.05--0.08 toward 0.04. In the later stage, once a smaller number of larger grains has formed, the weights become nearly stationary, both within each prediction segment and across segments, and the block ordering is preserved throughout. We emphasize that, under the scale assignment used here, [4,4,4,2,2,2,1,1,1], blocks 0--2 operate at the coarsest resolution (largest receptive field, largest-displacement motion), blocks 3--5 at an intermediate resolution, and blocks 6--8 at the finest, full resolution. The routing is therefore dominated throughout by a finest-scale block (block 6) together with intermediate-scale blocks (blocks 3, 4 and 5) and one coarse block (block 0), while the remaining blocks contribute little. This behavior is consistent with normal grain-growth kinematics, where inter-frame boundary motion is small and therefore well captured by fine-scale blocks. The early shift toward the finest-scale block reflects the transition from an initially fine-grained state toward smoother, statistically self-similar coarsening. The highly non-uniform routing weights show that the model learns specialized multiscale activations. We further observe that the routing weights vary smoothly across the phase-field correction segments, indicating that the learned scale combination tracks the physical microstructural state and is not disrupted by the corrective integration.

\subsection{Computation time cost comparison}

To evaluate the acceleration performance of the proposed framework, we profiled the inference and simulation scripts. The VFN was trained on the HPC platform, and the trained model was then downloaded for inference-time profiling. All tests were performed on an Ubuntu 24.04 desktop equipped with an Intel i9-13900KF CPU, 64 GB RAM, and an NVIDIA RTX A4000 GPU. We first measured the inference time of the VFN and found that generating 18 predicted snapshots required only about 0.87 s. By comparison, phase-field simulation of the same time interval (18,000 time steps), implemented with NVIDIA Numba-CUDA on the same GPU, required about 1332 s. This result indicates that the VFN provides an approximately three-orders-of-magnitude speedup. In addition, because the VFN does not exhibit the strong growth in computational cost associated with increasing microstructural length scale, its efficiency advantage over phase-field simulation is expected to become even greater for larger simulation domains.

\begin{figure}[pos=htbp]
    \centering
    \includegraphics[width=0.7\linewidth]{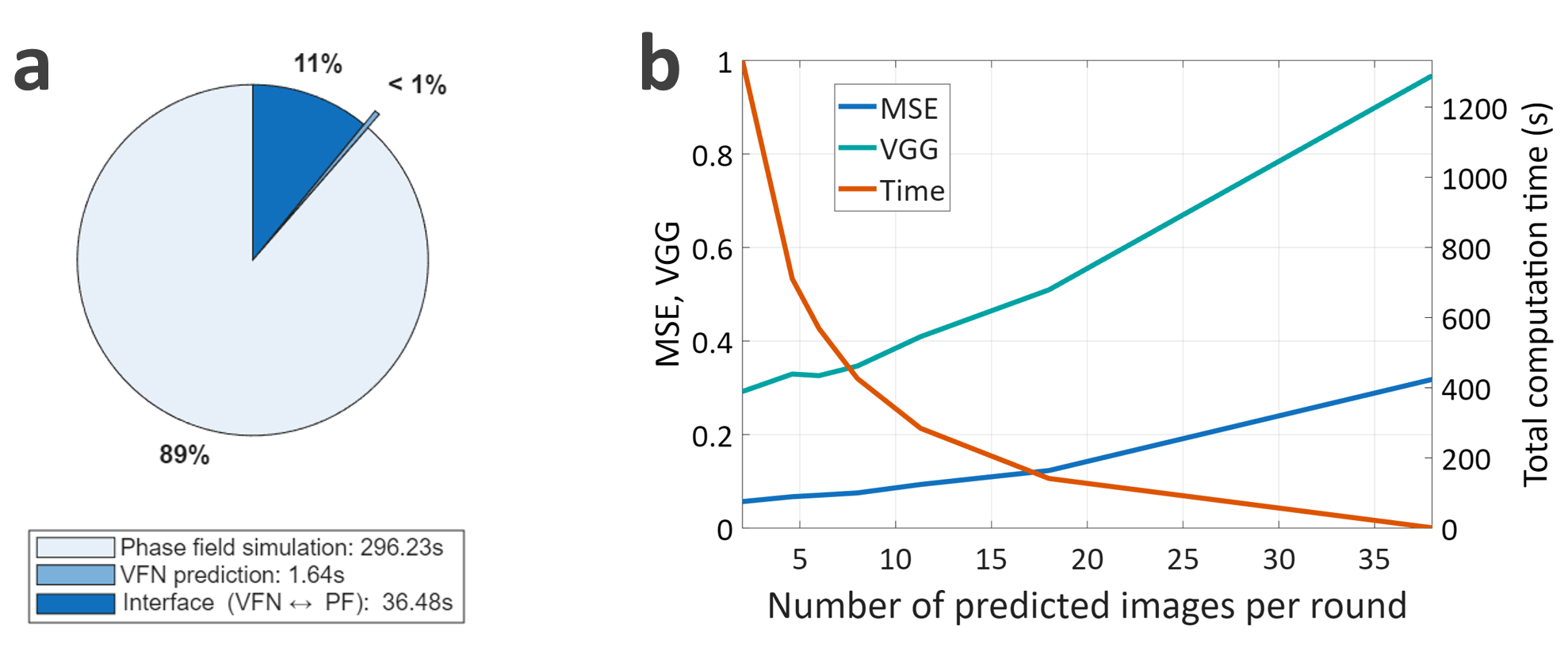}
    \caption{Computation time of the joint framework. (a) Computation time ratio of VFN prediction, PF simulation and interface between them for a joint prediction of 11-2-11-2-12 sequence as in figure \ref{fig:joint}(a); (b) Trade-off between computational cost and accuracy for varying numbers of VFN predictions per joint cycle. Increasing the number of VFN predictions per cycle reduces runtime but increases error, highlighting the balance between efficiency and precision.}
    \label{fig:time}
\end{figure}

We next profiled the joint framework, which consists primarily of VFN prediction, the interface between the VFN and PF solver, and PF simulation. In practice, the computational cost of VFN prediction and data transfer between the two modules is negligible compared with that of PF simulation, as shown in figure \ref{fig:time}(a). Consequently, the total runtime of the joint framework depends mainly on the fraction of snapshots generated by PF simulation. For the 38-snapshot case, inserting more PF steps reduces the number of snapshots predicted by the VFN, which increases computational cost but also improves accuracy by suppressing error accumulation. This trade-off between time cost and accuracy is shown in figure \ref{fig:time}(b), where computation time and error metrics are plotted as functions of the number of VFN-predicted frames per cycle for a fixed total of 38 frames. Here, each PF segment contains two simulated frames, and the x-axis denotes the number of consecutive frames predicted by the VFN between two PF correction segments. As the number of VFN-predicted frames per cycle increases, the number of PF insertions decreases. At $x=38$, corresponding to direct VFN prediction without PF correction, the computation time is minimized but the error is maximized. Conversely, shorter VFN segments improve accuracy at the expense of higher runtime. The best balance is achieved at approximately 10--18 VFN steps per cycle, where the framework maintains substantial speedup (about 5 times faster) over full PF simulation while significantly improving prediction accuracy. Overall, the joint scheme enables accurate long-horizon grain-growth prediction at a fraction of the computational cost of direct phase-field simulation, providing a practical approach for large-scale microstructure evolution modeling.

\subsection{Generalization across different PF models}

\begin{figure}[pos=htbp]
    \centering
    \includegraphics[width=0.8\linewidth]{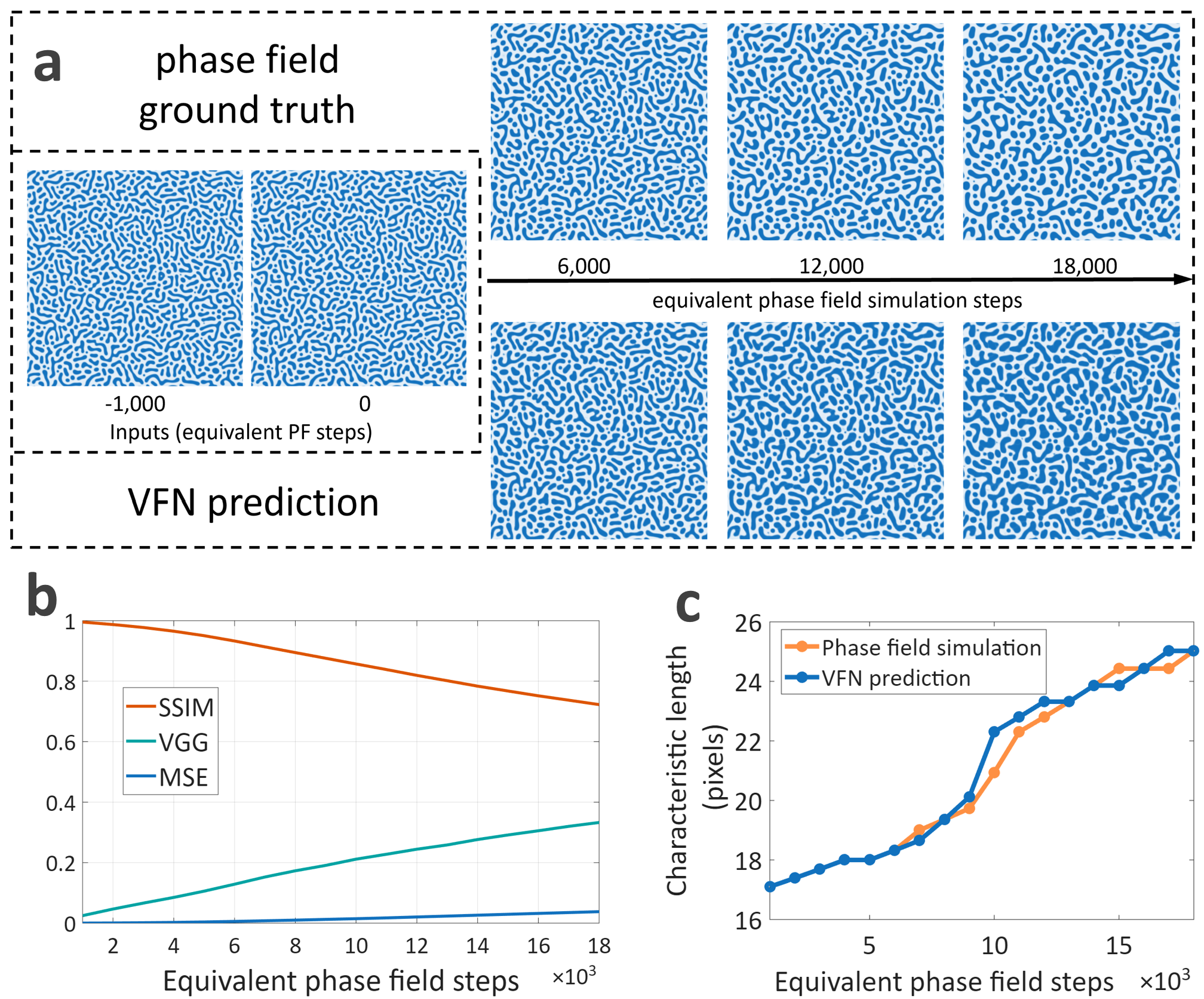}
    \caption{Voxel-flow network prediction results for spinodal decomposition. (a) Input images, ground-truth images, and predicted images of spinodal decomposition evolution (the full 18-frame sequence is shown in figure S4); (b) SSIM, MSE, and VGG loss curves; and (c) time evolution of the characteristic length of the spinodal pattern from VFN prediction and phase-field simulation.}
    \label{fig:vfnspinodal}
\end{figure}

To further examine the generalizability of the prediction method, we extended the analysis to another phase-field model: spinodal decomposition, originally introduced by Cahn. The formulation of this PF model is given in section \ref{sec:PF}. Random initial configurations were generated, and the VFN was trained using the same procedure as for the grain-growth case. The snapshot interval remained 1,000 time steps, and the image size was 512$\times$512. The results are presented in the same format as those for grain growth in figure \ref{fig:vfnspinodal}(a). The VFN successfully predicted the major phase-separation process, although some detailed differences remained compared with the phase-field ground truth. The coarsening of existing stripes and the disappearance of small phase regions were captured by the model, whereas some topological connections between adjacent phase regions were predicted less accurately. To compute quantitative error metrics, the pixel colors of the predicted images were mapped to concentration values in the range $(0,1)$ by interpolating RGB values from the colormap. The MSE was then calculated as:
\begin{equation}
    \mathrm{MSE} = \frac{1}{W \times H} \sum_{x=1}^{W} \sum_{y=1}^{H} \big(c_{\text{pred}}(x,y) - c_{\text{PF}}(x,y)\big)^2
\end{equation} 
VGG loss and SSIM were also computed following the same approach as for grain growth, with the results shown in figure \ref{fig:vfnspinodal}(b). For the final predicted image, the MSE, VGG perceptual loss, and SSIM were 3.79\%, 0.333, and 0.722, respectively. Owing to the diffuse nature of spinodal patterns, the MSE values were lower than those for grain growth. However, the SSIM scores were comparatively worse, reflecting the difficulty of accurately predicting fine-scale interfacial structures. In particular, the model occasionally failed to conserve the concentration ratio between the two phases, and some phase-boundary shapes were not faithfully reproduced. Nevertheless, the network successfully predicted the overall long-term morphology. We then applied a 2D FFT to identify the dominant peak in the isotropic frequency domain, corresponding to the characteristic wavelength. Importantly, this key physical descriptor—the characteristic length of the spinodal pattern—was consistent between predictions and simulations, as shown in figure \ref{fig:vfnspinodal}(c). Across the 18 predicted frames, the VFN reproduced the characteristic length exactly for the first six frames. The nonlinear increase observed between the 9th and 11th frames was also accurately captured, and the final predicted frame remained consistent with the ground truth. Overall, these results confirm that the voxel-flow network generalizes well across different types of phase-field models. The combination of visual agreement, quantitative metrics, and physical descriptors highlights the model's ability to preserve key microstructural features, even under challenging conditions.

\section{Conclusion}

The joint VFN-PF framework enables accurate and efficient long-horizon prediction of microstructure evolution. For direct VFN prediction, the MSE remains 6.72\% for grain growth and 3.79\% for spinodal decomposition after 18,000 phase-field steps, while key physical descriptors such as grain number, grain area distribution, and spinodal characteristic length are well preserved. By periodically inserting short phase-field correction segments, the joint framework suppresses error accumulation, improves physical consistency, and retains the flexibility and generalizability of an image-driven method across different microstructure evolution problems. Both direct VFN prediction and the joint strategy provide substantial computational acceleration, and the joint framework further extends statistically reliable prediction to 82,000 phase-field steps.

Future work will focus on developing a prediction module that can better adapt to different model parameters and further improve the generalizability of the framework. It will also be important to validate the framework on real experimental microstructures, extend it to more complex multi-physics phase-field systems with chemical, mechanical, or thermal coupling, and introduce adaptive correction strategies that dynamically determine when phase-field simulation should be reinserted to optimize both efficiency and accuracy.

\section{Data availability}

The data and code for this work are available at Zhou, Leeroy (2026), ``Joint prediction framework on microstructure evolution,'' Mendeley Data, V1, doi: 10.17632/ngtshs7k22.1.

\section{Acknowledgments} 

The authors would like to acknowledge the funding by the Science and Technology Commission of Shanghai Municipality (25DZ3001902, 23TS1401600). The computations in this paper were run on the Siyuan-1 and Zhiyuan-1 clusters supported by the Center for High Performance Computing at Shanghai Jiao Tong University.

\section{Competing Interests}

All authors declare no financial or non-financial competing interests. 

\section{Declaration of Generative AI and AI-assisted technologies in the writing process}

During the preparation of this work, the authors used Prism to improve language and readability and to assist with \LaTeX{} formatting. After using this tool, the authors reviewed and edited the content as needed and take full responsibility for the content of the published article.

\printcredits

\bibliographystyle{unsrtnat}
\bibliography{ref_abbr.bib}

@article{razavi2026physics,
  title={Physics-informed GCN-LSTM framework for long-term forecasting of 2D and 3D microstructure evolution},
  author={Razavi, Hamidreza and Moelans, Nele},
  journal={npj Comput. Mater.},
  year={2026},
}

@article{RN100,
   author = {Shi, Xingjian and Chen, Zhourong and Wang, Hao and Yeung, Dit-Yan and Wong, Wai-Kin and Woo, Wang-chun},
   title = {Convolutional LSTM network: A machine learning approach for precipitation nowcasting},
   journal = {Adv. Neural Inf. Process. Syst.},
   volume = {28},
   year = {2015},
   type = {Journal Article}
}

@inproceedings{RN99,
   author = {Gao, Zhangyang and Tan, Cheng and Wu, Lirong and Li, Stan Z},
   title = {Simvp: Simpler yet better video prediction},
   booktitle = {Proc. IEEE/CVF Conf. Comput. Vis. Pattern Recognit.},
   pages = {3170–3180},
   type = {Conference Proceedings},
   year = {2022}
}

@inproceedings{tan2023openstl,
  title={OpenSTL: A Comprehensive Benchmark of Spatio-Temporal Predictive Learning},
  author={Tan, Cheng and Li, Siyuan and Gao, Zhangyang and Guan, Wenfei and Wang, Zedong and Liu, Zicheng and Wu, Lirong and Li, Stan Z},
  booktitle={NeurIPS Datasets Benchmarks Track},
  year={2023}
}

@article{RN97,
   author = {Tourret, Damien and Liu, Hong and LLorca, Javier},
   title = {Phase-field modeling of microstructure evolution: Recent applications, perspectives and challenges},
   journal = {Prog. Mater. Sci.},
   volume = {123},
   pages = {100810},
   year = {2022},
   type = {Journal Article}
}

@article{RN96,
   author = {Shen, Zhong-Hui and Wang, Jian-Jun and Jiang, Jian-Yong and Huang, Sharon X and Lin, Yuan-Hua and Nan, Ce-Wen and Chen, Long-Qing and Shen, Yang},
   title = {Phase-field modeling and machine learning of electric-thermal-mechanical breakdown of polymer-based dielectrics},
   journal = {Nat. Commun.},
   volume = {10},
   number = {1},
   pages = {1843},
   year = {2019},
   type = {Journal Article}
}

@article{RN95,
   author = {Steinbach, Ingo},
   title = {Phase-field models in materials science},
   journal = {Modell. Simul. Mater. Sci. Eng.},
   volume = {17},
   number = {7},
   pages = {073001},
   year = {2009},
   type = {Journal Article}
}

@article{oommen2022learning,
  title={Learning two-phase microstructure evolution using neural operators and autoencoder architectures},
  author={Oommen, Vivek and Shukla, Khemraj and Goswami, Somdatta and Dingreville, R{\'e}mi and Karniadakis, George Em},
  journal={npj Comput. Mater.},
  volume={8},
  number={1},
  pages={190},
  year={2022},
  publisher={Nature Publishing Group UK London}
}

@article{lanzoni2024extreme,
  title={Extreme time extrapolation capabilities and thermodynamic consistency of physics-inspired neural networks for the 3D microstructure evolution of materials via Cahn--Hilliard flow},
  author={Lanzoni, Daniele and Fantasia, Andrea and Bergamaschini, Roberto and Pierre-Louis, Olivier and Montalenti, Francesco},
  journal={Mach. Learn.: Sci. Technol.},
  volume={5},
  number={4},
  pages={045017},
  year={2024},
  publisher={IOP Publishing}
}

@article{farizhandi2023spatiotemporal,
  title={Spatiotemporal prediction of microstructure evolution with predictive recurrent neural network},
  author={Farizhandi, Amir Abbas Kazemzadeh and Mamivand, Mahmood},
  journal={Comput. Mater. Sci.},
  volume={223},
  pages={112110},
  year={2023},
  publisher={Elsevier}
}

@article{zhu2024spatiotemporal,
  title={Spatiotemporal evolution of grain microstructure: A CNN perspective},
  author={Zhu, Changsheng and Liu, Shuo and Gao, Zihao and Wang, Lijun and Miao, Jintao},
  journal={Mater. Today Commun.},
  volume={40},
  pages={110005},
  year={2024},
  publisher={Elsevier}
}

@article{yang2021self,
  title={Self-supervised learning and prediction of microstructure evolution with convolutional recurrent neural networks},
  author={Yang, Kaiqi and Cao, Yifan and Zhang, Youtian and Fan, Shaoxun and Tang, Ming and Aberg, Daniel and Sadigh, Babak and Zhou, Fei},
  journal={Patterns},
  volume={2},
  number={5},
  pages={100243},
  year={2021},
  publisher={Elsevier}
}

@article{fan2024accelerate,
  title={Accelerate microstructure evolution simulation using graph neural networks with adaptive spatiotemporal resolution},
  author={Fan, Shaoxun and Hitt, Andrew L and Tang, Ming and Sadigh, Babak and Zhou, Fei},
  journal={Mach. Learn.: Sci. Technol.},
  volume={5},
  number={2},
  pages={025027},
  year={2024},
  publisher={IOP Publishing}
}

@article{qin2024graingnn,
  title={GrainGNN: A dynamic graph neural network for predicting 3D grain microstructure},
  author={Qin, Yigong and DeWitt, Stephen and Radhakrishnan, Balasubramaniam and Biros, George},
  journal={J. Comput. Phys.},
  volume={510},
  pages={113061},
  year={2024},
  publisher={Elsevier}
}

@article{Zapiain2021accelerating,
Author = {Zapiain, David Montes de Oca and Stewart, James A. and Dingreville, Remi},
Title = {Accelerating phase-field-based microstructure evolution predictions via
   surrogate models trained by machine learning methods},
Journal = {npj Comput. Mater.},
Year = {2021},
Volume = {7},
Number = {1},
pages = {3},
}

@article{Hu2022accelerating,
Author = {Hu, C. and Martin, S. and Dingreville, R.},
Title = {Accelerating phase-field predictions via recurrent neural networks
   learning the microstructure evolution in latent space},
Journal = {Comput. Methods Appl. Mech. Eng.},
Year = {2022},
Volume = {397},
pages={115128}
}

@article{wight2020solving,
  title={Solving Allen-Cahn and Cahn-Hilliard equations using the adaptive physics informed neural networks},
  author={Colby L, Wight and Zhao, Jia},
  journal={Commun. Comput. Phys.},
  volume={29},
  issue={3},
  pages={930–954},
  year={2021}
}

@article{ciesielski2025deep,
  title={Deep operator network surrogate for phase-field modeling of metal grain growth during solidification},
  author={Ciesielski, Danielle and Li, Yulan and Hu, Shenyang and King, Ethan and Corbey, Jordan and Stinis, Panos},
  journal={Comput. Mater. Sci.},
  volume={246},
  pages={113417},
  year={2025},
  publisher={Elsevier}
}

@article{tseng2023deep,
  title={Deep learning model to predict ice crystal growth},
  author={Tseng, Bor-Yann and Guo, Chen-Wei Conan and Chien, Yu-Chen and Wang, Jyn-Ping and Yu, Chi-Hua},
  journal={Adv. Sci.},
  volume={10},
  number={21},
  pages={2207731},
  year={2023},
  publisher={Wiley Online Library}
}

@article{chen2024mau,
  title={L-MAU: A multivariate time-series network for predicting the Cahn-Hilliard microstructure evolutions via low-dimensional approaches},
  author={Chen, Sheng-Jer and Yu, Hsiu-Yu},
  journal={Comput. Phys. Commun.},
  volume={305},
  pages={109342},
  year={2024},
  publisher={Elsevier}
}

@article{rieger2024setting,
  title={Setting the standard for machine learning in phase field prediction: a benchmark dataset and baseline metrics},
  author={Rieger, Laura Hannemose and Zeli{\v{c}}, Klemen and Mele, Igor and Katra{\v{s}}nik, Toma{\v{z}} and Bhowmik, Arghya},
  journal={Sci. Data},
  volume={11},
  number={1},
  pages={1275},
  year={2024},
  publisher={Nature Publishing Group UK London}
}

@article{peivaste2022machine,
  title={Machine-learning-based surrogate modeling of microstructure evolution using phase-field},
  author={Peivaste, Iman and Siboni, Nima H and Alahyarizadeh, Ghasem and Ghaderi, Reza and Svendsen, Bob and Raabe, Dierk and Mianroodi, Jaber Rezaei},
  journal={Comput. Mater. Sci.},
  volume={214},
  pages={111750},
  year={2022},
  publisher={Elsevier}
}

@article{wu2025simgate,
  title={SimGate: A deep learning surrogate model for predicting microstructure evolution using the phase-field method},
  author={Wu, Pin and Huang, Haiwang and Yang, Qingcheng and Qian, Bo and Gao, Yongxin and Yang, Yiguo and Zhang, Huiran and Zhen, Qiang},
  journal={Comput. Mater. Sci.},
  volume={256},
  pages={113883},
  year={2025},
  publisher={Elsevier}
}

@article{yan2022novel,
  title={A novel physics-regularized interpretable machine learning model for grain growth},
  author={Yan, Weishi and Melville, Joseph and Yadav, Vishal and Everett, Kristien and Yang, Lin and Kesler, Michael S and Krause, Amanda R and Tonks, Michael R and Harley, Joel B},
  journal={Mater. Des.},
  volume={222},
  pages={111032},
  year={2022},
  publisher={Elsevier}
}

@article{wang2022multi,
  title={Multi-input convolutional network for ultrafast simulation of field evolvement},
  author={Wang, Zhuo and Yang, Wenhua and Xiang, Linyan and Wang, Xiao and Zhao, Yingjie and Xiao, Yaohong and Liu, Pengwei and Liu, Yucheng and Banu, Mihaela and Zikanov, Oleg and others},
  journal={Patterns},
  volume={3},
  number={6},
  pages={100494},
  year={2022},
  publisher={Elsevier}
}

@InProceedings{hu2023dynamic,
    author    = {Hu, Xiaotao and Huang, Zhewei and Huang, Ailin and Xu, Jun and Zhou, Shuchang},
    title     = {A Dynamic Multi-Scale Voxel Flow Network for Video Prediction},
    booktitle = {Proc. IEEE/CVF Conf. Comput. Vis. Pattern Recognit. (CVPR)},
    year      = {2023},
    pages     = {6121-6131}
}

@article{cahn1961spinodal,
  title={On spinodal decomposition},
  author={Cahn, John W},
  journal={Acta Metall.},
  volume={9},
  number={9},
  pages={795--801},
  year={1961},
  publisher={Elsevier}
}

@article{chen1994computer,
  title={Computer simulation of the domain dynamics of a quenched system with a large number of nonconserved order parameters: The grain-growth kinetics},
  author={Chen, Long-Qing and Yang, Wei},
  journal={Phys. Rev. B},
  volume={50},
  number={21},
  pages={15752},
  year={1994},
  publisher={APS}
}

@article{PINNoriginal,
title = {Physics-informed neural networks: A deep learning framework for solving forward and inverse problems involving nonlinear partial differential equations},
journal = {J. Comput. Phys.},
volume = {378},
pages = {686-707},
year = {2019},
author = {M. Raissi and P. Perdikaris and G.E. Karniadakis},
}

@InProceedings{perceptualloss,
author="Johnson, Justin
and Alahi, Alexandre
and Fei-Fei, Li",
editor="Leibe, Bastian
and Matas, Jiri
and Sebe, Nicu
and Welling, Max",
title="Perceptual Losses for Real-Time Style Transfer and Super-Resolution",
booktitle="Computer Vision -- ECCV 2016",
year="2016",
publisher="Springer International Publishing",
address="Cham",
pages="694-711",
}

@ARTICLE{wang2004image,
  author={Zhou Wang and Bovik, A.C. and Sheikh, H.R. and Simoncelli, E.P.},
  journal={IEEE Trans. Image Process.}, 
  title={Image quality assessment: from error visibility to structural similarity}, 
  year={2004},
  volume={13},
  number={4},
  pages={600-612},
}

@article{scikit-image,
 title = {scikit-image: image processing in {P}ython},
 author = {van der Walt, {S}t\'efan and {S}ch\"onberger, {J}ohannes {L}. and
           {Nunez-Iglesias}, {J}uan and {B}oulogne, {F}ran\c{c}ois and {W}arner,
           {J}oshua {D}. and {Y}ager, {N}eil and {G}ouillart, {E}mmanuelle and
           {Y}u, {T}ony and the scikit-image contributors},
 year = {2014},
 volume = {2},
 pages = {e453},
 journal = {PeerJ},
}

@inproceedings{paszke2019pytorch,
  title={PyTorch: An Imperative Style, High-Performance Deep Learning Library},
  author={Paszke, Adam and Gross, Sam and Massa, Francisco and Lerer, Adam and Bradbury, James and Chanan, Gregory and Killeen, Trevor and Lin, Zeming and Gimelshein, Natalia and Antiga, Luca and Desmaison, Alban and Kopf, Andreas and Yang, Edward and DeVito, Zachary and Raison, Martin and Tejani, Alykhan and Chilamkurthy, Sasank and Steiner, Benoit and Fang, Lu and Bai, Junjie and Chintala, Soumith},
  booktitle={Adv. Neural Inf. Process. Syst. 32},
  year={2019}
}

@article{harris2020array,
  title={Array programming with {NumPy}},
  author={Harris, Charles R and Millman, K Jarrod and van der Walt, St{\'e}fan J and Gommers, Ralf and Virtanen, Pauli and Cournapeau, David and Wieser, Eric and Taylor, Julian and Berg, Sebastian and Smith, Nathaniel J and Kern, Robert and Picus, Matti and Hoyer, Stephan and van Kerkwijk, Marten H and Brett, Matthew and Haldane, Allan and del R{\'\i}o, Jaime Fern{\'a}ndez and Wiebe, Mark and Peterson, Pearu and G{\'e}rard-Marchant, Pierre and Sheppard, Kevin and Reddy, Tyler and Weckesser, Warren and Abbasi, Hameer and Gohlke, Christoph and Oliphant, Travis E},
  journal={Nature},
  volume={585},
  number={7825},
  pages={357-362},
  year={2020},
  publisher={Nature Publishing Group},
}

@Inbook{Haridasan2025,
author="Haridasan, Navaneeth
and Krishnaveni, V. S.
and Sandra, S.
and Abhijith, M. S.",
editor="Sachin Kumar, S.
and Ashok, Neelesh
and Sukumar, N.
and Mohan, Neethu
and Soman, K. P.
and Thomas, Sabu",
title="Physics Informed Neural Networks: Fundamentals and Application to Phase Field Models",
bookTitle="Artificial Intelligence for Materials Informatics",
year="2025",
publisher="Springer Nature Switzerland",
address="Cham",
pages="103-153",
}

@article{pinnphaseflow,
    author = {Qiu, Rundi and Huang, Renfang and Xiao, Yao and Wang, Jingzhu and Zhang, Zhen and Yue, Jieshun and Zeng, Zhong and Wang, Yiwei},
    title = {Physics-informed neural networks for phase-field method in two-phase flow},
    journal = {Phys. Fluids},
    volume = {34},
    number = {5},
    pages = {052109},
    year = {2022},
}

@INPROCEEDINGS{pinnhardconstraint,
  author={Chen, Pengyu and Jiang, Benben},
  booktitle={Proc. China Autom. Congr. (CAC)}, 
  title={Physics-Informed Neural Networks for Learning Allen-Cahn Equation Using Hard Constraint Representation}, 
  year={2024},
  volume={},
  number={},
  pages={6433-6438},
}

@article{pfpinnschen,
title = {PF-PINNs: Physics-informed neural networks for solving coupled Allen-Cahn and Cahn-Hilliard phase field equations},
journal = {J. Comput. Phys.},
volume = {529},
pages = {113843},
year = {2025},
author = {Nanxi Chen and Sergio Lucarini and Rujin Ma and Airong Chen and Chuanjie Cui},
}

@article{pinnsmpf,
title = {PINNs-MPF: A Physics-Informed Neural Network framework for Multi-Phase-Field simulation of interface dynamics},
journal = {Eng. Anal. Bound. Elem.},
volume = {176},
pages = {106200},
year = {2025},
author = {Seifallah Elfetni and Reza Darvishi Kamachali},
}

@article{deeponetbazant,
title = {Phase-Field DeepONet: Physics-informed deep operator neural network for fast simulations of pattern formation governed by gradient flows of free-energy functionals},
journal = {Comput. Methods Appl. Mech. Eng.},
volume = {416},
pages = {116299},
year = {2023},
author = {Wei Li and Martin Z. Bazant and Juner Zhu},
}

@misc{gangmei2025,
      title={Learning coupled Allen-Cahn and Cahn-Hilliard phase-field equations using Physics-informed neural operator(PINO)}, 
      author={Gaijinliu Gangmei and Santu Rana and Bernard Rolfe and Kishalay Mitra and Saswata Bhattacharyya},
      year={2025},
      eprint={2507.18731},
      archivePrefix={arXiv},
      primaryClass={cs.CE},
}

@article{regularization,
   author = {Zhu, H. X. and Thorpe, S. M. and Windle, A. H.},
   title = {The geometrical properties of irregular two-dimensional Voronoi tessellations},
   journal = {Philos. Mag. A},
   volume = {81},
   number = {12},
   pages = {2765-2783},
   year = {2007},
   type = {Journal Article}
}

@article{RN92,
   author = {Hötzer, Johannes and Seiz, Marco and Kellner, Michael and Rheinheimer, Wolfgang and Nestler, Britta},
   title = {Phase-field simulation of solid state sintering},
   journal = {Acta Mater.},
   volume = {164},
   pages = {184-195},
   year = {2019},
   type = {Journal Article}
}

@article{RN93,
   author = {Chen, Lei and Zhang, Hao Wei and Liang, Lin Yun and Liu, Zhe and Qi, Yue and Lu, Peng and Chen, James and Chen, Long-Qing},
   title = {Modulation of dendritic patterns during electrodeposition: A nonlinear phase-field model},
   journal = {J. Power Sources},
   volume = {300},
   pages = {376-385},
   year = {2015},
   type = {Journal Article}
}

@article{RN94,
   author = {Shibuta, Yasushi and Okajima, Yoshinao and Suzuki, Toshio},
   title = {Phase-field modeling for electrodeposition process},
   journal = {Sci. Technol. Adv. Mater.},
   volume = {8},
   number = {6},
   pages = {511},
   year = {2007},
   type = {Journal Article}
}

\end{document}